\documentclass[aps,prd,twocolumn,groupedaddress,showpacs]{revtex4-1}
\usepackage{epsf,epsfig,graphicx}
\usepackage{amsmath}
\usepackage{amssymb}
\usepackage{tikz}

\def\edth{\;\raise1.0pt\hbox{$'$}\hskip-6pt\partial\;}
\def\baredth{\;\overline{\raise1.0pt\hbox{$'$}\hskip-6pt
\partial}\;}
\def\gsim{~\rlap{$>$}{\lower 1.0ex\hbox{$\sim$}}}

\newcommand{\be}{\begin{equation}}
\newcommand{\ee}{\end{equation}}
\newcommand{\bw}{\begin{widetext}}
\newcommand{\ew}{\end{widetext}}

\newcommand{\intinf}{\int_{-\infty}^{\infty}}
\newcommand{\suml}{\sum_{l=0}^{\infty}}
\newcommand{\summ}{\sum_{m=-\ell}^{\ell}}

\usepackage{xcolor}

\usepackage[colorlinks]{hyperref}
\definecolor{darkblue}{HTML}{2E3092}
\hypersetup{colorlinks=true,linkcolor=darkblue,citecolor=darkblue,urlcolor=darkblue}

\AtBeginDocument{\renewcommand{\d}{\mathbf{d}}}

\begin{document}

\title{Overlap reduction functions for a polarized stochastic gravitational-wave
background in the Einstein Telescope-Cosmic Explorer and the LISA-Taiji networks}

\author{Guo-Chin Liu$^{1}$ and Kin-Wang Ng$^{2,3}$}

\affiliation{
$^1$Department of Physics, Tamkang University, Tamsui, New Taipei City 25137, Taiwan\\
$^2$Institute of Physics, Academia Sinica, Taipei 11529, Taiwan\\
$^3$Institute of Astronomy and Astrophysics, Academia Sinica, Taipei 11529, Taiwan
}

\vspace*{0.6 cm}
\date{\today}
\vspace*{1.2 cm}

\begin{abstract}
The detection of gravitational waves from the coalescences of binary compact stars by current
interferometry experiments has opened up a new era of gravitational-wave astrophysics and cosmology.
The search for a stochastic gravitational-wave background is underway by correlating signals from a pair of
detectors in the detector network formed by the LIGO, Virgo, and KAGRA.
In a previous work, we have developed a method based on spherical harmonic expansion to calculate the overlap
reduction functions of the LIGO-Virgo-KAGRA network for a polarized stochastic gravitational-wave background. 
In this work, we will apply the method to calculate the overlap reduction functions of third-generation detectors such as a ground-based network linking the Einstein Telescope, the Cosmic Explorer, and the LISA-Taiji joint space mission.
\end{abstract}

\maketitle

\section{Introduction}

The successful detection of gravitational waves (GWs) from compact binary coalescences in the LIGO-Virgo-KAGRA collaboration~\cite{LVK2021} has accelerated the implementation of third-generation detectors such as the ground-based Einstein Telescope~\cite{et2010} and Cosmic Explorer~\cite{ce2019}, as well as the space-borne LISA~\cite{lisa2017}, DECIGO~\cite{decigo2011}, Taiji~\cite{taiji2017}, and Tianqin~\cite{tianqin2016}.

Stochastic gravitational-wave background (SGWB) is one of the main goals in GW experiments~\cite{romano}. 
Many astrophysical and cosmological sources for SGWB at various frequency ranges have been proposed, 
including distant merging compact binaries, early phase transitions, cosmic string or domain wall networks, 
second-order density perturbations, and inflationary GWs. 
These relic GWs provide a powerful probe into the production mechanisms deep in the early Universe;
however, they still remain elusive in detection.
It is indeed a big challenge to observation because the background level is highly uncertain, 
dependent on the origin of the sources as well as the frequency ranges.
Recently, using the data from Advanced LIGO's and Advanced Virgo's third observing run (O3) combined with the earlier O1 and O2 runs, upper limits have been derived on an isotropic SGWB  
with the spectral energy density of $\Omega_{\rm GW}(f)\le 3.4\times 10^{-9}$ 
at a reference frequency of $f=25\,{\rm Hz}$
for a $f^\alpha$ power-law spectrum with a spectral index of $\alpha=2/3$ and $\Omega_{\rm GW}\le 5.8\times 10^{-9}$  for a frequency independent SGWB~\cite{ligo2101}.

In general, the SGWB can be anisotropic and polarized. There have been a lot of works that provide mechanisms to generate intensity and polarization anisotropies in the SGWB. A net circular polarization of the SGWB is predicted when 
helical GWs are produced in specific axion inflation models involving axion-gauge-field interactions~\cite{alexander,satoh,sorbo,crowder2013}. Intensity anisotropies of the SGWB has been recently explored~\cite{bartolo,pitrou}. Furthermore, linear polarization can be generated through multiple gravitational Compton scattering processes off of massive compact objects during the GW propagation across the large scale structures of the Universe~\cite{cusin,pizzuti}. 
The Advanced LIGO and Advanced Virgo has further placed an upper limit on an anisotropic SGWB, 
ranging from $\Omega_{\rm GW}(f)< (0.57-9.3)\times 10^{-9} {\rm sr}^{-1}$ at $f=25\,{\rm Hz}$, 
depending on observational direction and spectral index~\cite{ligo2103}. 
Recently, there has been a search for a non-zero circular polarization in the LIGO-Virgo O3 data, concluding that there is no preference for helical versus non-helical GWs in the SGWB~\cite{martinovic21}.

The current method in GW interferometry experiments for detecting SGWB is to correlate the responses of a pair of detectors with a fixed baseline to the GW strain amplitude. 
This allows us to reduce uncorrelated detector noises and obtain a large GW signal~\cite{romano}. 
The correlation is a mapping of the SGWB sky with the overlap reduction functions (ORFs) of the detector pair~\cite{michelson1987,christensen1992,flanagan1993,allen1997,allen1999,cornish2001,seto2006,seto2007,seto2008,thrane2009,crowder2013,romano2017}, thus encoding the intensity and polarization anisotropies or the Stokes parameters of the SGWB in the correlation data. In our previous work~\cite{chu21}, we have provided a method based on spherical harmonic expansion of the polarization basis tensors, from which a fast numerical algorithm is developed to compute ORF multipoles of the detector pair with respect to the SGWB Stokes parameters. This fast computation of the ORFs will be necessary for constructing data pipelines to extract the anisotropy and polarization power spectra in future SGWB measurements.  In this paper, we will apply this method to calculate the ORF multipoles of the detector pairs formed by the  LISA and the Taiji space missions, as well as the Einstein Telescope and the Cosmic Explorer (ET-CE) ground-based detectors.

The paper is organized as follows. The next section introduces SGWB, whose correlated signals and ORFs in interferometers are recapitulated in Secs.~\ref{CorrSignal} and \ref{ORFsection}, respectively. The long-wavelength limit for the response of an equilateral triangle of interferometers will be discussed in Sec.~\ref{triangle}. 
In Secs.~\ref{ETCEpair} and ~\ref{LTpair}, we will work out the ORF multipoles for the Einstein Telescope-Cosmic Explorer network and the LISA-Taiji network, respectively. Section.~\ref{conclusion} is our conclusion.

\section{Polarized SGWB}

In the Minkowskian spacetime $(t,\vec{x})$, the metric perturbation $h_{ij}$ in the transverse traceless gauge depicts traveling GWs at the speed of light $c=\omega/k$. It can be expanded by Fourier modes as
\be
\label{eq:planwave}
h_{ij}(t,\vec{x}) = \sum_{A}\intinf \d f \int_{S^2} \d\hat{k} \;
    h_A(f,\hat{k}) \mathbf{e}^A_{ij}(\hat{k})
    e^{-2 \pi i f (t - \hat{k}\cdot \vec{x}/c)}\,,
\ee
where $A$ stands for the polarization of GWs with the basis tensors $\mathbf{e}^A_{ij}(\hat{k})$, which are transverse to the propagation direction, $\hat{k}$. Here $h_{ij}$ is treated as real, so the Fourier components with negative frequencies are given by $h_A(-f,\hat{k})=h_A^*(f,\hat{k})$ for all $f\ge 0$. We define a SGWB as a collection of GWs satisfying the condition that $h_{ij}$ are random Gaussian fields with a statistical behavior completely characterized by
the two-point correlation function $\langle h_{ij}(t,\vec{x}_1) h_{ij}(t,\vec{x}_2) \rangle$, where the angle brackets denote
their ensemble averages. The ensemble averages of the Fourier modes have the following form				
\be
\label{eq:paa}
\langle h_{A}(f,\hat{k}) h^*_{A'}(f',\hat{k}') \rangle
= \delta(f-f') \delta(\hat{k}-\hat{k}')P_{AA'}(f,\hat{k}) \,,
\ee
where the spatial translational invariance dictates the delta function of their 3-momenta, $\delta(\vec{k}-\vec{k}')$. Note that the power spectra $P_{AA'}(f,\hat{k})$ remain to be direction dependent.

For GWs coming from the sky direction $-\hat{k}$ with wave vector $\vec{k}$, it is customary to write the polarization basis tensors in terms of the basis vectors in the spherical coordinates,
\begin{align}
\label{basisvector}
    \mathbf{e}^{+}(\hat{k}) &= \hat{\mathbf{e}}_\theta \otimes \hat{\mathbf{e}}_\theta  
                               - \hat{\mathbf{e}}_\phi \otimes \hat{\mathbf{e}}_\phi \,, \nonumber \\
    \mathbf{e}^{\times}(\hat{k}) &= \hat{\mathbf{e}}_\theta \otimes \hat{\mathbf{e}}_\phi  
                                    + \hat{\mathbf{e}}_\phi \otimes \hat{\mathbf{e}}_\theta  \,,
\end{align}
in which $\hat{\mathbf{e}}_\theta$, $\hat{\mathbf{e}}_\phi$, and $\hat{k}$ form a right-handed orthonormal basis.
Also, we can define the complex circular polarization basis tensors as
\begin{align}
    \mathbf{e}_{R} &= \frac{(\mathbf{e}_{+} + i \mathbf{e}_{\times})}{\sqrt{2}} \,,
   &\mathbf{e}_{L} &= \frac{(\mathbf{e}_{+} - i \mathbf{e}_{\times})}{\sqrt{2}} \,,
\end{align}
where $\mathbf{e}_{R}$ stands for the right-handed GW with a positive helicity while $\mathbf{e}_{L}$ stands for the left-handed GW with a negative helicity. The corresponding amplitudes in Eq.~(\ref{eq:planwave}) in the two different bases are related to each other via
\begin{align}
    h_{R} &= \frac{(h_{+} - i h_{\times})}{\sqrt{2}} \,,
   &h_{L} &= \frac{(h_{+} + i h_{\times})}{\sqrt{2}} \,.
\end{align}

The coherency matrix $P_{AA'}$ in Eq.~(\ref{eq:paa}) is related to the Stokes parameters, $I$, $Q$, $U$, and $V$ as 
\begin{align}
    I &= \left[ \langle h_R h_R^* \rangle + \langle h_L h_L^* \rangle \right] / 2 \,,\nonumber\\
    Q + iU &=  \langle h_L h_R^*  \rangle \,,\nonumber\\
    Q - iU &=  \langle h_R h_L^*  \rangle \,,\nonumber\\
    V &= \left[ \langle h_R h_R^* \rangle - \langle h_L h_L^* \rangle \right] / 2 \,\label{eq:spIQUV},
\end{align}
which are functions of the frequency $f$ and the propagation direction $\hat{k}$.
$I$ is the intensity, $Q$ and $U$ represent the linear polarization, and $V$ is the circular polarization.

\section{Signal Correlation}
\label{CorrSignal}

The signal $h_a(t_a,\vec{x}_a)$ in a $90^\circ$ GW interferometer $a$, such as LIGO, Virgo, or KAGRA, located at $\vec{x}_a$ can be expressed as the contraction of the metric perturbation $h_{ij}(t,\vec{x})$ and the detector tensor $d_a^{ij}$ of the detector:
\bw
\begin{align}
    h_a(t_a,\vec{x}_a) 
    &= d_a^{ij} h_{ij}(t_a,\vec{x}_a) \nonumber \\
    &= 
        d_a^{ij}
        \sum_{A}\intinf \d f \int_{S^2} \d\hat{k} \;
        h_A(f,\hat{k}) \mathbf{e}^A_{ij}(\hat{k})
        e^{-2 \pi i f (t_a - \hat{k}\cdot \vec{x}_a/c)} \,,
\end{align}
\ew
where the detector tensor is
\begin{align}
    d_a^{ij} &= 
    \frac{1}{2}
    \left(
    \mathbf{X}_a^{i}
    \mathbf{X}_a^{j}-
    \mathbf{Y}_a^{i}
    \mathbf{Y}_a^{j}
    \right) \,,
\end{align}
with $\mathbf{X}_a^{i}$ being the $i$-th component of the unit vector along the X arm of the detector, 
while $\mathbf{Y}_a^{i}$ represents the Y arm.

\begin{figure}
\centering
\includegraphics[width=0.4\textwidth]{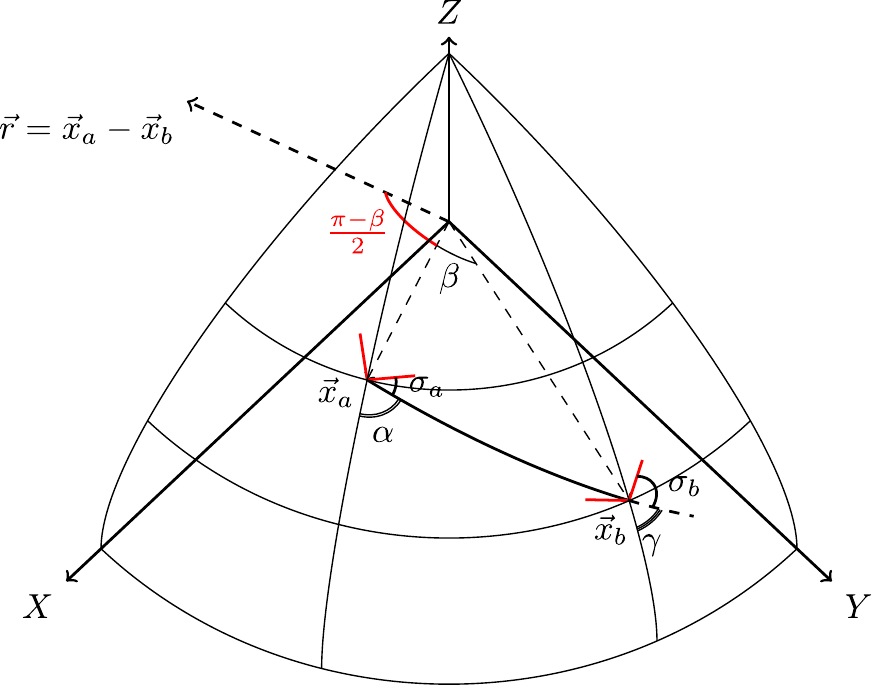}
    \caption[Coordinate]{Convention of angles. $\vec{x}_a$, $\vec{x}_b$ represent the positions of detector $a$ and detector $b$, respectively. $\vec{r}$ is the baseline. $\sigma_a$ and $\sigma_b$ are the angles between the great circle connecting the pair $a$-$b$ and the X arms of detector $a$ and detector $b$, respectively.}
\label{fig:angles}
\end{figure}

The correlation of signals in a pair of detectors $a$ and $b$ can be expressed in terms of the baseline vector $\vec{r}\equiv\vec{x}_a - \vec{x}_b$ and the time delay $\tau \equiv t_a-t_b$. In frequency domain, we have
\bw
\begin{align}
   \xi_{ab}( f, \vec{r} ) 
  =& 
    \int_{-T/2}^{T/2} \d \tau \;
    \langle h_a(t_a,\vec{x}_a) h_b^*(t_b,\vec{x}_b) \rangle
    \; e^{2\pi i f \tau} 
    \nonumber \\
  =&
   \, d_a^{ij} d_b^{kl}
    \int_{S^2} \d \hat{k} \;
    \sum_{AA'}
    P_{AA'}(f,\hat{k})
    \mathbf{e}^{A}_{ij}(\hat{k}) \mathbf{e}^{*A'}_{kl}(\hat{k})
    e^{2\pi i f (\hat{k}\cdot\vec{r}/c)} 
   \nonumber \\
  =&
    \label{eq:xi2gamma}
    \sum_{S=\{I,V,Q\pm iU\}}
    \int_{S^2} \d \hat{k} \;
    S(f,\hat{k}) \;
    {}^{ijkl}\mathbb{D}_{ab} \;
    \mathbb{E}^{S}_{ijkl} (\hat{k}) 
    e^{2\pi i f (\hat{k}\cdot\vec{r}/c)} \,,
\end{align}
\ew
where the Fourier integral is taken over an interval $T$ within which the orientation and the condition of the detectors are approximately fixed. 
In addition, the interval $T$ has to be large enough when compared with the period of GW signals in the detectors.
In Eq.~(\ref{eq:xi2gamma}), the polarization tensors $\mathbb{E}^S$ associated with the corresponding Stokes parameters are defined as
\begin{align}
    \mathbb{E}^{I}_{ijkl} (\hat{k}) &= 
    \mathbf{e}^{R}_{ij}(\hat{k}) \mathbf{e}^{*R}_{kl}(\hat{k}) + 
    \mathbf{e}^{L}_{ij}(\hat{k}) \mathbf{e}^{*L}_{kl}(\hat{k}) \,, \nonumber
    \\ 
    \mathbb{E}^{V}_{ijkl} (\hat{k}) &= 
    \mathbf{e}^{R}_{ij}(\hat{k}) \mathbf{e}^{*R}_{kl}(\hat{k}) - 
    \mathbf{e}^{L}_{ij}(\hat{k}) \mathbf{e}^{*L}_{kl}(\hat{k}) \,, \nonumber
    \\ 
    \mathbb{E}^{Q+iU}_{ijkl} (\hat{k}) &= 
    \mathbf{e}^{L}_{ij}(\hat{k}) \mathbf{e}^{*R}_{kl}(\hat{k}) \,, \nonumber
    \\ 
    \label{eq:EEQ-iU}
    \mathbb{E}^{Q-iU}_{ijkl} (\hat{k}) &= 
    \mathbf{e}^{R}_{ij}(\hat{k}) \mathbf{e}^{*L}_{kl}(\hat{k}) \,,
\end{align}
while $\mathbb{D}_{ab}$ denotes the direct product of two detector tensors
\begin{align}
    {}^{ijkl}\mathbb{D}_{ab}
    &= 
    d^{ij}_a(\theta_a, \phi_a, \sigma_a) 
    d^{kl}_b(\theta_b, \phi_b, \sigma_b)
\,,
\end{align}
where the coordinate angles are illustrated in Fig.~\ref{fig:angles}.  We further define the ORFs as
\begin{align}
    \label{eq:gamma}
    \gamma^{S}(\hat{k},f,\vec{r})
    = \mathbb{D}
    \cdot
    \mathbb{E}^S(\hat{k}) 
    e^{2\pi i f (\hat{k}\cdot\vec{r}/c)} \,.
\end{align}

In our previous work~\cite{chu21}, we compute the correlation~(\ref{eq:xi2gamma}) in the spherical harmonic basis:
\begin{align}
   \label{eq:xilm} 
   \xi_{ab}(f)
   &=
    \sum_{S=\{I,V,Q\pm iU\}}
   \sum_{\ell m}  S_{\ell m}(f) \gamma_{\ell m}^{S}(f) \,,
\end{align}
where we have expanded the Stokes parameters in terms of ordinary and spin-weighted spherical harmonics as
\begin{align}
    I(f,\hat{k}) &= \sum_{\ell m}I_{\ell m}(f) \; Y_{\ell m}(\hat{k}) \,,\nonumber\\
    V(f,\hat{k}) &= \sum_{\ell m}V_{\ell m}(f) \; Y_{\ell m}(\hat{k}) \,,\nonumber\\
    (Q+iU)(f,\hat{k}) &= \sum_{\ell m}(Q+iU)_{\ell m}(f) \; _{+4}Y_{\ell m}(\hat{k}) \,,\nonumber\\
    \label{eq:Q-iUlm}
    (Q-iU)(f,\hat{k}) &= \sum_{\ell m}(Q-iU)_{\ell m}(f) \; _{-4}Y_{\ell m}(\hat{k}) \,,
\end{align}
so as the ORFs:
\begin{align}
    \label{eq:gammaIVlm}
    \gamma_{\ell m}^{I,V}(f) 
    &=\int_{S^2} \d \hat{k} \; Y_{\ell m}(\hat{k}) \gamma^{I,V}(\hat{k},f) \,,\\
    \label{eq:gammaQUlm}
    \gamma_{\ell m}^{Q\pm i U}(f) 
    &=\int_{S^2} \d \hat{k} \; _{\pm4}Y_{\ell m}(\hat{k}) \gamma^{Q\pm iU}(\hat{k},f)\,.
\end{align}

By plugging Eq.~(\ref{eq:gamma}) into Eqs.~(\ref{eq:gammaIVlm}) and~(\ref{eq:gammaQUlm}), expanding the polarization basis tensors as
\begin{align}
    \mathbb{E}^{I}_{ijkl} (\hat{k}) 
    &=\sum_{\ell_e m_e} {}_{ijkl}\mathbb{E}^{I}_{\ell_e m_e} Y_{\ell_e m_e}(\hat{k}) \,, \nonumber
   \\
    \mathbb{E}^{V}_{ijkl} (\hat{k}) 
    &=\sum_{\ell_e m_e} {}_{ijkl}\mathbb{E}^{V}_{\ell_e m_e} Y_{\ell_e m_e}(\hat{k}) \,, \nonumber
   \\
    \mathbb{E}^{Q+iU}_{ijkl} (\hat{k}) 
    &=\sum_{\ell_e m_e} {}_{ijkl}\mathbb{E}^{Q+iU}_{\ell_e m_e} {}_{-4}Y_{\ell_e m_e}(\hat{k}) \,, \nonumber
   \\
    \label{eq:EEQ-iUlm}
    \mathbb{E}^{Q-iU}_{ijkl} (\hat{k}) 
    &=\sum_{\ell_e m_e} {}_{ijkl}\mathbb{E}^{Q-iU}_{\ell_e m_e} {}_{+4}Y_{\ell_e m_e}(\hat{k}) \,,
\end{align}
and using the spherical wave expansion:
\be
e^{i\vec{k}\cdot\vec{r}} = 4\pi \suml \summ i^\ell j_\ell(kr) Y_{\ell m}^*(\hat{k}) Y_{\ell m}(\hat{r}) \,,
\ee
where $j_\ell(x)$ is the spherical Bessel function, we obtain the ORF multipoles as
\bw
\begin{align}
    \label{eq:gammaIV_lm}
    \gamma_{\ell m}^{I,V}(f) 
  &=
    (4\pi) 
    \sum_{m'}
    D^{\ell}_{m' m}(-\alpha,-\theta_a,-\phi_a) 
    \sum_{\ell_e m_e}
    \mathbb{D}_0
    \cdot
    \mathbb{E}^{I,V}_{\ell_e m_e} 
    \sum_{L M} 
    i^L j_L\left(\frac{2\pi f r}{c}\right) 
    Y_{L M}(\hat{r}_0)  
    \left\langle
    \begin{matrix}
        L  && \ell_e   && \ell \\
        M  && 0\; m_e  && 0\;  m'
    \end{matrix}
    \right\rangle 
    \,, \\
    \label{eq:gammaQU_lm}
    \gamma_{\ell m}^{Q\pm iU}(f) 
  &=
    (4\pi) 
    \sum_{m'}
    D^{\ell}_{m' m}(-\alpha,-\theta_a,-\phi_a) 
    \sum_{\ell_e m_e}
    \mathbb{D}_0
    \cdot
    \mathbb{E}^{Q\pm iU}_{\ell_e m_e} 
    \sum_{L M} 
    i^L j_L\left(\frac{2\pi f r}{c}\right) 
    Y_{L M}(\hat{r}_0)  
    \left\langle
    \begin{matrix}
        L  && \ell_e        && \ell \\
        M  && \mp4\; m_e    && \pm4\;  m'
    \end{matrix}
    \right\rangle \,,
\end{align}
where we have used the shorthand notation for the integral of three spherical harmonics given by
\begin{align}
    \left\langle
    \begin{matrix}
        L && l_1  &&  l_2 \\
        M && s_1\;m_1  && s_2\;m_2 
    \end{matrix}
    \right\rangle
    &\equiv
    \int \d \hat{k} \;
    Y_{LM}^*(\hat{k})\; {}_{s_1}\!Y_{l_1 m_1}(\hat{k}) \; {}_{s_2}\!Y_{l_2 m_2}(\hat{k})
    \nonumber \\
    &=
    \label{eq:threeJ}
    (-1)^M
    \sqrt{\frac{(2L+1)(2l_1+1)(2l_2+1)}{4\pi}}
    \begin{pmatrix}
        L && l_1  &&  l_2 \\
        0 && -s_1  &&  -s_2 
    \end{pmatrix}
    \begin{pmatrix}
        L && l_1  &&  l_2 \\
       -M && m_1  &&  m_2 
    \end{pmatrix} \,,
\end{align}
\ew
which involves two Wigner-3j symbols~\cite{book:Varshalovich}. In Eqs.~(\ref{eq:gammaIV_lm}) and~(\ref{eq:gammaQU_lm}), $\mathbb{D}_0\cdot \mathbb{E}^S$ are the antenna pattern functions evaluated in a coordinate system by placing the detector $a$ at the north pole of the Earth and the detector $b$ on the $\phi=0$ meridian. They are listed in Appendix~\ref{sec:DE}. In this coordinate system, the corresponding baseline direction 
$\hat{r}_0$ is  
\be
\label{eq:r0}
\hat{r}_0 = (\theta_{r_0},\phi_{r_0})=(\frac{\beta-\pi}{2}, 0)=(\frac{\pi-\beta}{2}, \pi)\,.
\ee
The Wigner-D matrices $D^{\ell}_{m' m}(-\alpha,-\theta_a,-\phi_a)$, where $(\theta_a,\phi_a)$ is the spherical coordinates of the detector $a$ and $\alpha$ is the angle between its meridian and the great circle connecting the pair $a$-$b$ as shown in Fig.~\ref{fig:angles}, account for the rotation of the whole pair of detectors to their actual positions.
They are related to the spin-weighted spherical harmonics by
\begin{align}
    \label{eq:DtoSWSH}
    D^{\ell}_{s m}(\alpha,\beta,\gamma)
    = \sqrt{\frac{4\pi}{2\ell+1}} {}_{-s}Y_{\ell m}(-\beta,-\gamma) e^{-i s \alpha} \,.
\end{align}
Also, the ORF multipoles have the conjugate relations:
\begin{align}
\gamma_{\ell -m}^{I,V}=&(-1)^{\ell+m}\gamma_{\ell m}^{I,V*}\,, \label{IVconjugate}\\
\gamma_{\ell -m}^{Q\pm iU}=&(-1)^{\ell+m}\gamma_{\ell m}^{Q\pm iU*}\,. \label{QUconjugate}
\end{align}

\bw
\section{Overlap Reduction Functions}
\label{ORFsection}

\subsection{Isotropic SGWB}

For an isotropic SGWB, the only relevant ORFs are $\gamma_{00}^{I}$ and $\gamma_{00}^{V}$. From Eq.~(\ref{eq:gammaIV_lm}) and Appendix~\ref{sec:DE}, we reproduce the known analytic results found in Ref.~\cite{seto2008}:
\begin{align}
    \gamma_{00}^{I}(f) 
    =&
    \sqrt{4\pi}
    \sum_{\ell_e m_e}
    \mathbb{D}_0
    \cdot
    \mathbb{E}^{I}_{\ell_e m_e} 
    i^{\ell_e} j_{\ell_e}\left(\frac{2\pi f r}{c}\right) 
    Y_{\ell_e m_e}(\hat{r}_0)  \nonumber \\
    =&
    \cos (2 (\sigma_1-\sigma_2)) \,C_1(f)   + \cos (2 (\sigma_1+\sigma_2)+\pi ) \,C_2(f) \,, \label{eq:gammaI_iso}\\
    \gamma_{00}^{V}(f) =&
      \sqrt{4\pi} 
    \sum_{\ell_e m_e}
    \mathbb{D}_0
    \cdot
    \mathbb{E}^{V}_{\ell_e m_e} 
    i^{\ell_e} j_{\ell_e}\left(\frac{2\pi f r}{c}\right) 
    Y_{\ell_e m_e}(\hat{r}_0)  \nonumber \\
 =&
    -\sin (2 (\sigma_1+\sigma_2)+\pi )\,C_3(f) \,, \label{eq:gammaV_iso}
\end{align}
which do not depend on the orientation of the detector pair as expected for an isotropic source, and where

\begin{align}
    C_1(f)
    =&
    \frac{4\sqrt{\pi}}{5}
    \left[\left(j_0+\frac{5 j_2}{7}+\frac{3 j_4}{112}\right) \cos ^4\left(\frac{\beta }{2}\right) \right] \,,
   \nonumber \\
   C_2(f)
   =& \frac{4\sqrt{\pi}}{5}  
    \left[
      \left(\frac{-3 j_0}{8} +\frac{45 j_2}{56}-\frac{169 j_4}{896}\right)
      +\left(\frac{j_0}{2}-\frac{5 j_2}{7}-\frac{27 j_4}{224}\right) \cos (\beta ) 
      +\left(-\frac{j_0}{8}-\frac{5 j_2}{56}-\frac{3 j_4}{896}\right) \cos (2 \beta )
    \right] \,, 
     \nonumber  \\
   C_3(f)
    =& 
    \frac{4\sqrt{\pi}}{5} \sin\left(\frac{\beta }{2}\right)
     \left[   \left(-j_1+\frac{7 j_3}{8}\right)
      +\left(j_1+\frac{3 j_3}{8}\right) \cos (\beta ) 
    \right] \nonumber \,.
\end{align} 
\ew

\subsection{SGWB anisotropy and polarization}

Numerical results of the ORF multipole moments for $l\le 4$ in Eqs.~(\ref{eq:gammaIV_lm}) and~(\ref{eq:gammaQU_lm}) for the pair formed by LIGO-Hanford and LIGO-Livingston
are obtained in Ref.~\cite{chu21}, 
where $\gamma_{\ell m}^{I}$'s and $\gamma_{00}^{V}$ match those made in Ref.~\cite{romano2017}, 
and the ORF multipoles for circular and linear polarizations are also found.

\begin{figure}
\centering
\includegraphics[width=0.6\textwidth]{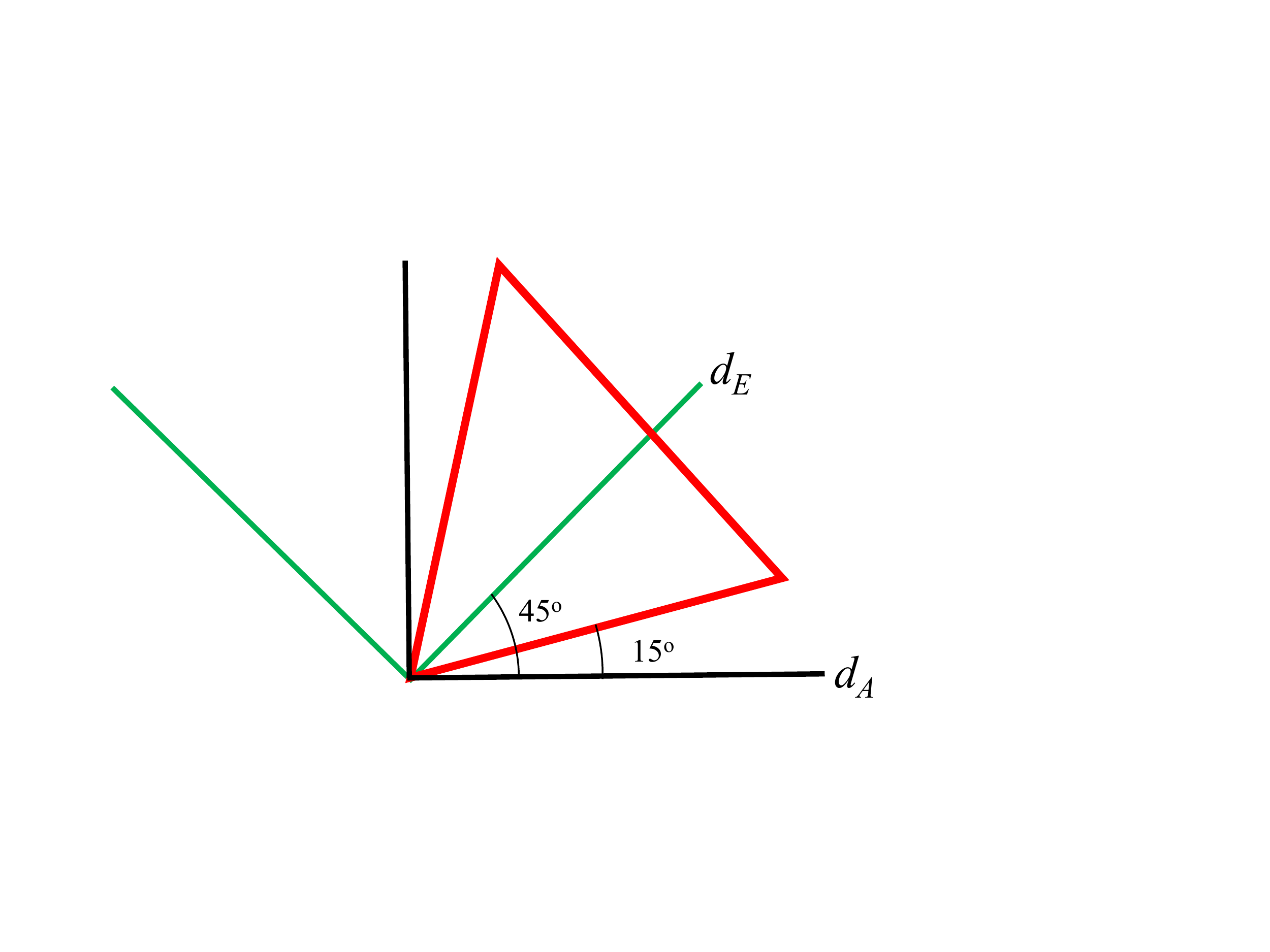}
    \caption[Coordinate]{An equilateral triangular GW detector. It behaves as two $90^\circ$ interferometers in the low frequency limit, represented by $d_A$ and $d_E$.}
\label{fig:et}
\end{figure}

\section{Equilateral Triangular Detector}
\label{triangle}

Third-generation GW detectors such as the Einstein Telescope~\cite{et2010} and LISA~\cite{lisa2017} 
have three arms in an equilateral triangular shape. 
This allows the detectors to resolve both GW polarizations and have better detector noise characterization. 
At each vertex of the triangle, a Michelson interferometer is placed to measure the difference between 
the time taken for a laser light to go a round trip along one interferometer arm and that along the other arm. 

In an ideal situation, the covariance noise matrix of the three interferometers is symmetric and thus can be diagonalized to form three orthogonal $A$, $E$, and $T$ channels~\cite{adams2010,smith2019}. In the low frequency limit ($fd/c\ll 1$), where $d$ is the arm length, the $T$ channel is suppressed and only the $A$ and $E$ channels are sensitive to GW signals. These two channels behave as two $90^\circ$ interferometers separated by $45^\circ$ with detector tensors $d^{ij}_A$ and $d^{ij}_E$, respectively~\cite{hild2010,mentasti2021}. Figure~\ref{fig:et} shows an equilateral triangular GW detector and the orientation of $d_A$ and $d_E$ relative to the triangle.

In this paper, we study the observation of the SGWB made by the ET-CT and the LISA-Taiji detector pairs. The separation distance $r$ between the pair detectors is typically much larger than the arm lengths of the detectors. In the following, we consider GWs in the frequency range that satisfies the low frequency limit, $fd/c\ll 1$,
but $fr/c$ is not necessarily small. This allows us to treat each equilateral triangular detector as $90^\circ$ interferometers and use the well established method to correlate the signals from the detector pairs.

\section{Einstein Telescope-Cosmic Explorer pair}
\label{ETCEpair}

The Einstein Telescope has equilateral triangular interferometers of $10\,{\rm km}$ arm length. The two underground candidate sites for the ET are under consideration: one in Sardinia and one at the Euregio Meuse-Rhine. The Cosmic Explorer has two sites in the United States, one $40\,{\rm km}$ in length and a second $20\,{\rm km}$ in length, each with a $90^\circ$ interferometer. The increase in arm length in addition to new detector technologies will greatly improve the sensitivity and bandwidth of the instruments.

The actual locations and arm orientations for the ET and the CE are yet to be determined. Suppose the separation distance between ET and CE be approximately given by $r\simeq 7500\, {\rm km}$; then the separation angle will be $\beta\simeq 36^\circ$. For GW frequencies $f\ll 3\times 10^4 {\rm Hz}$,
this network will provide us with two distinct correlation signals: the CE's $90^\circ$ interferometer with the ET's $d_A$ and $d_E$ detector tensors, respectively. The former has the antenna pattern functions $\mathbb{D}_0\cdot \mathbb{E}^S$ listed in Appendix~\ref{sec:DE} with 
$\sigma_1=\sigma_{{\rm ET},A}$ and $\sigma_2=\sigma_{\rm CE}$. 
The $\sigma_{{\rm ET},A}$ is the angle between the great circle connecting ET-CE and the X arm of the ET's $d_A$, 
which is $15^\circ$ shifted clockwise from one side of the ET triangle as shown in Fig.~\ref{fig:et}. 
The $\sigma_{\rm CE}$ is the angle between the great circle connecting ET-CE and the X arm of the CE interferometer in either site. The latter has $\sigma_1=\sigma_{{\rm ET},E}$ and $\sigma_2=\sigma_{\rm CE}$, 
where $\sigma_{{\rm ET},E}=\sigma_{{\rm ET},A}+45^\circ$. 

Lastly, to compute the ORF multipoles in Eqs.~(\ref{eq:gammaIV_lm}) and~(\ref{eq:gammaQU_lm}), one needs the Wigner-D matrices $D^{\ell}_{m' m}(-\alpha,-\theta_a,-\phi_a)$, where $(\theta_a,\phi_a)$ is the spherical coordinates of the ET and $\alpha$ is the angle between its meridian and the great circle connecting ET-CE.
Once all the configurational angles are determined, the ORF multipoles can be readily produced using the same algorithm developed in Ref.~\cite{chu21}. Since ORFs are oscillatory functions of the product $fr$, except some phase differences due to different configurational angles, the results for the ORF multipole moments get similar to those for the LIGO's Hanford-Livingston pair (whose separation distance is $3002\,{\rm km}$) obtained in Ref.~\cite{chu21} by simply rescaling the frequency by a factor of $3002/7500$. Here we are therefore not to explicitly show the ET-CE ORFs by assuming some arbitrary configurational angles.

\begin{figure}
\centering
\includegraphics[width=0.45\textwidth]{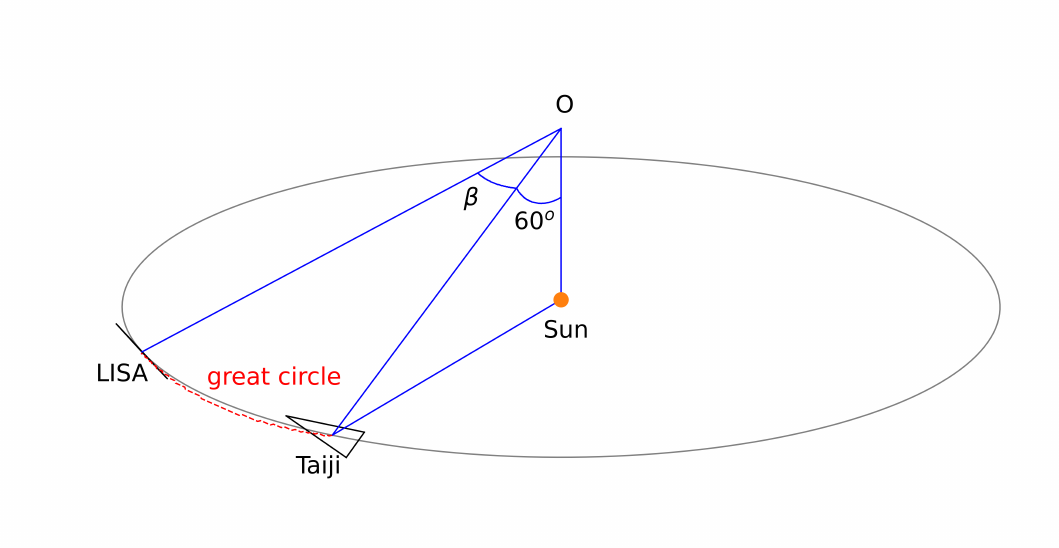}
    \caption[Coordinate]{The LISA-Taiji network.}
\label{fig:LisaTaiji}
\end{figure}

\begin{figure}
\centering
\includegraphics[width=0.5\textwidth]{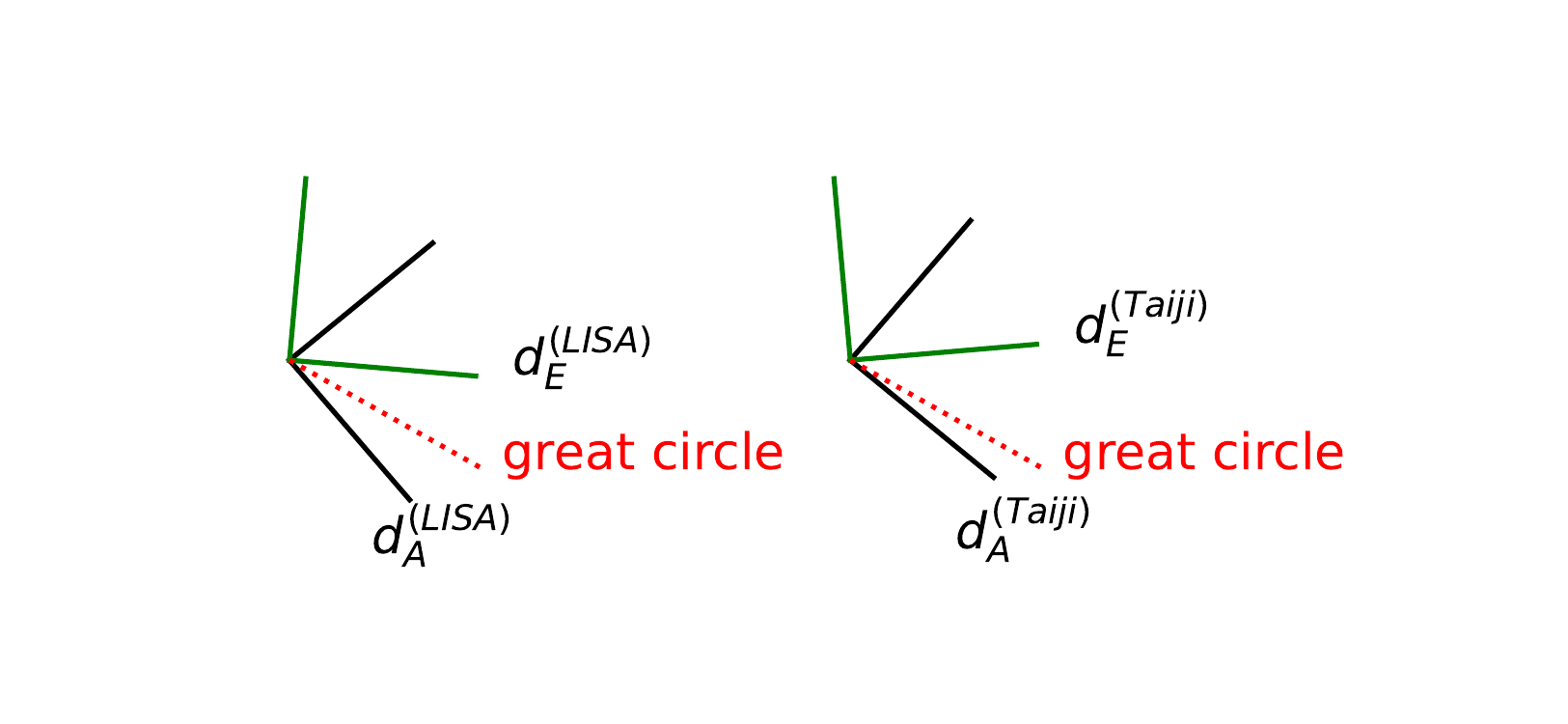}
    \caption[Coordinate]{The LISA-Taiji detector tensors.}
\label{fig:detectortensor}
\end{figure}

\section{LISA-Taiji pair}
\label{LTpair}

LISA has three spacecrafts located at the corners of an equilateral triangle in space, with each side $d=2.5\times 10^6\, {\rm km}$, trailing behind the Earth by $20^\circ$ in a heliocentric orbit. The detector plane is inclined to the ecliptic plane by $30^\circ$. The triangle is spinning clockwise as viewed from the Sun, around its center and synchronous with the Earth's orbit with a period of $1\,{\rm yr}$. Taiji's mission configuration is similar to LISA's, except that it moves $20^\circ$ ahead of the Earth with longer arm length $d=3\times 10^6\,{\rm km}$.  Their separation distance is given by $r=2R_E \sin 20^\circ \simeq 0.68\,{\rm AU}$, where $R_E=1\,{\rm AU}=1.5\times 10^8\,{\rm km}$. As a result, both detector planes are tangential to the surface of a sphere centered at $O$ of radius $R=R_E/\cos 30^\circ \simeq 1.55\,{\rm AU}$, and their separation angle is 
$\beta=2\sin^{-1}[r/(2R)]\simeq 34.5^\circ$. See Fig.~\ref{fig:LisaTaiji} for the LISA-Taiji network.

Figure.~\ref{fig:detectortensor} shows the $d_A$ and $d_E$ detector tensors of the LISA-Taiji pair in the low frequency limit, $f\ll 0.1 {\rm Hz}$. There are four correlations: $A$-$A$, $A$-$E$, $E$-$A$, and $E$-$E$, where $A$-$A$ denotes the correlation between the LISA's $d_A$ and the Taiji's $d_A$, and so on. 
Let the angle between the great circle connecting LISA-Taiji and the X arm of the LISA's $d_A$ be $\phi(t)$. Then, the angle between the great circle connecting LISA-Taiji and the X arm of the Taiji's $d_A$ will be given by $\phi(t)-\phi_0$, 
where $\phi_0$ is a constant phase angle difference yet to be fixed by the mission design. Note that $\phi(t)$ is a time dependent angle with a spinning period of $1\,{\rm yr}$. 
In Table~\ref{table:angles}, we summarize the assignment of the angles of 
the antenna pattern functions $\mathbb{D}_0\cdot \mathbb{E}^S$ in Appendix~\ref{sec:DE} for the four correlations.

\begin{table}
    \begin{tabular}{@{\quad}c@{\quad}l@{\quad}l@{\quad}l@{\quad}l@{\quad}}
    \hline
    \hline
        Correlation  &$\sigma_1$  &$\sigma_2$  &$\sigma_1-\sigma_2$  &$\sigma_1+\sigma_2$  \\
    \hline                                                       
        $A$-$A$     &$\phi$        &$\phi-\phi_0$                      &$\phi_0$                      &$2\phi-\phi_0$      \\
        $A$-$E$     &$\phi$        &$\phi-\phi_0+{\pi\over4}$    &$\phi_0-{\pi\over4}$    &$2\phi-\phi_0+{\pi\over4}$   \\
        $E$-$A$     &$\phi+{\pi\over4}$        &$\phi-\phi_0$    &$\phi_0+{\pi\over4}$    &$2\phi-\phi_0+{\pi\over4}$   \\
        $E$-$E$     &$\phi+{\pi\over4}$        &$\phi-\phi_0+{\pi\over4}$   &$\phi_0$    &$2\phi-\phi_0+{\pi\over2}$    \\

    \hline
    \hline
    \end{tabular}
    \caption{Angles in the antenna pattern functions for the LISA-Taiji correlation. $\sigma_{1,2}$ are the angles     between the great circle connecting LISA-Taiji and the X arms of the LISA's and Taiji's detector tensors, respectively. $A$ and $E$ denote the $A$ and $E$ channels, respectively.}
    \label{table:angles}
\end{table}

\subsection{Isotropic SGWB}

From Eqs.~(\ref{eq:gammaI_iso}) and~(\ref{eq:gammaV_iso}), we have
\begin{align}
    \gamma_{00,AA}^{I}(f) 
    =&
    \cos (2\phi_0) \,C_1(f)   - \cos (4\phi-2\phi_0) \,C_2(f) \,, \nonumber \\
   \gamma_{00,AE}^{I}(f) 
    =&
    \sin (2\phi_0) \,C_1(f)   + \sin (4\phi-2\phi_0) \,C_2(f) \,, \nonumber \\
\gamma_{00,EA}^{I}(f) 
    =&
    -\sin (2\phi_0) \,C_1(f)   + \sin (4\phi-2\phi_0) \,C_2(f) \,, \nonumber \\
\gamma_{00,EE}^{I}(f) 
    =&
    \cos (2\phi_0) \,C_1(f)   + \cos (4\phi-2\phi_0) \,C_2(f) \,,
\label{eq:gammaI_iso2}
\end{align}
and
\begin{align}
    \gamma_{00,AA}^{V}(f) =&-\gamma_{00,EE}^{V}(f) =
    \sin (4\phi-2\phi_0)\,C_3(f) \,, \nonumber \\
 \gamma_{00,AE}^{V}(f) =&  \gamma_{00,EA}^{V}(f) =
    \cos (4\phi-2\phi_0)\,C_3(f) \,.
\label{eq:gammaV_iso2}
\end{align}
Note that in general $\gamma_{00,AE}^{I}\neq \gamma_{00,EA}^{I}$. 

Generally, we can take the sum of $A$-$A$ and $E$-$E$ correlations to obtain from Eq.~(\ref{eq:xilm}) that
\begin{align}
   \xi_{ab}(f)
   &= 2 I_{00}  \cos (2\phi_0) \,C_1(f),
\end{align}
which is sensitive to $I_{00}$ only. Furthermore, the sum of $A$-$E$ and $E$-$A$ correlations gives
\begin{align}
  \xi_{ab}(f) =&  2 I_{00} \sin (4\phi-2\phi_0) \,C_2(f) + \nonumber \\
  & 2 V_{00} \cos (4\phi-2\phi_0)\,C_3(f)\,,
\end{align}
where the $C_2$ and $C_3$ terms are orthogonal to each other, so $I_{00}$ and $V_{00}$ can be extracted from $\xi_{ab}$ by sine and cosine transforms in the time-varying $\phi(t)$, respectively.

\subsection{SGWB anisotropy and polarization}
\label{ORFmultipole}

\begin{figure}[ht!]
\centering
\includegraphics[width=1.0\textwidth]{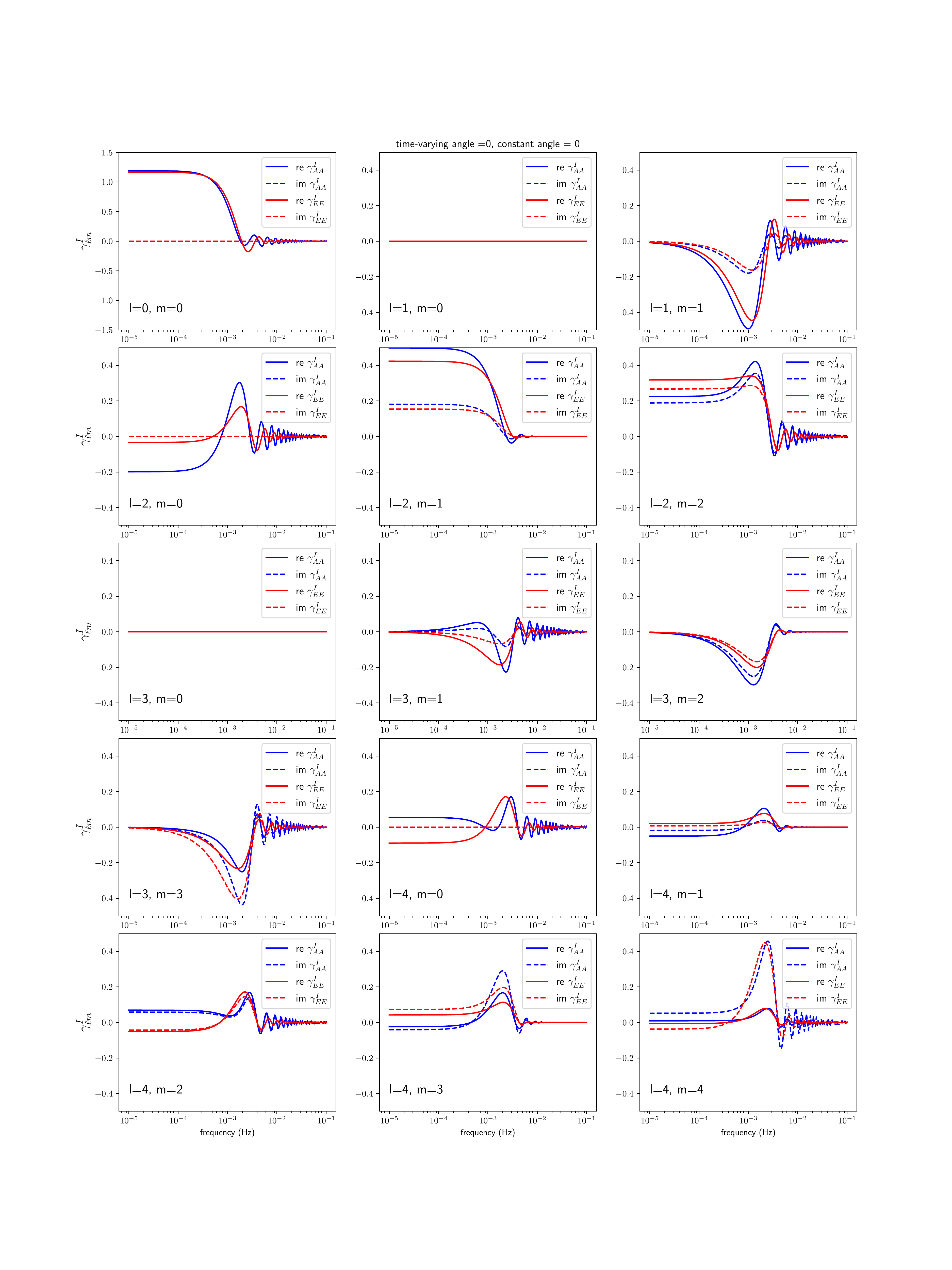}
    \caption[Coordinate]{Real and imaginary parts of the multipole moments of the intensity $A$-$A$ and $E$-$E$ overlap reduction function
    $\gamma_{\ell m}^I$ for the LISA-Taiji detector pair when $\phi(t)=0^\circ$ and the constant angle $\phi_0=0^\circ$.
    Plots of $\ell=0,1,2,3,4$ and $m\ge 0$ are shown. The $m<0$ multipoles can be obtained by using the conjugate relation in Eq.~(\ref{IVconjugate}).}
\label{fig:ORFI0AAEE}
\index{figures}
\end{figure}

\begin{figure}[ht!]
	\centering
	\includegraphics[width=1.0 \textwidth]{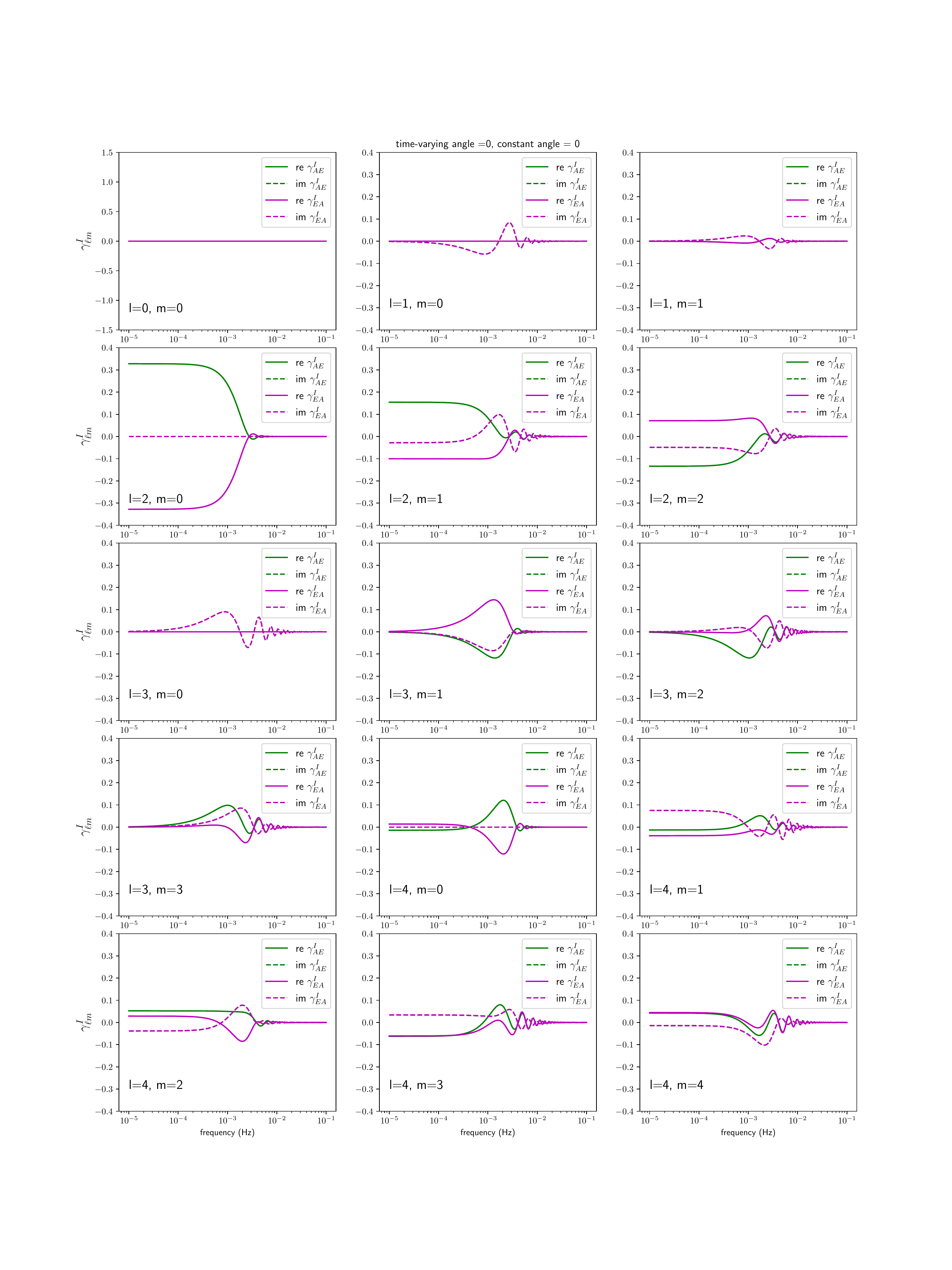}
	\caption[Coordinate]{Real and imaginary parts of the multipole moments of the intensity  $A$-$E$ and $E$-$A$ overlap reduction function
		$\gamma_{\ell m}^I$ for the LISA-Taiji detector pair when $\phi(t)=0^\circ$ and the constant angle $\phi_0=0^\circ$.
		Plots of $\ell=0,1,2,3,4$ and $m\ge 0$ are shown. The $m<0$ multipoles can be obtained by using the conjugate relation in Eq.~(\ref{IVconjugate}).}
	\label{fig:ORFI0AEEA}
	\index{figures}
\end{figure}

\begin{figure}[ht!]
\centering
\includegraphics[width=1.0\textwidth]{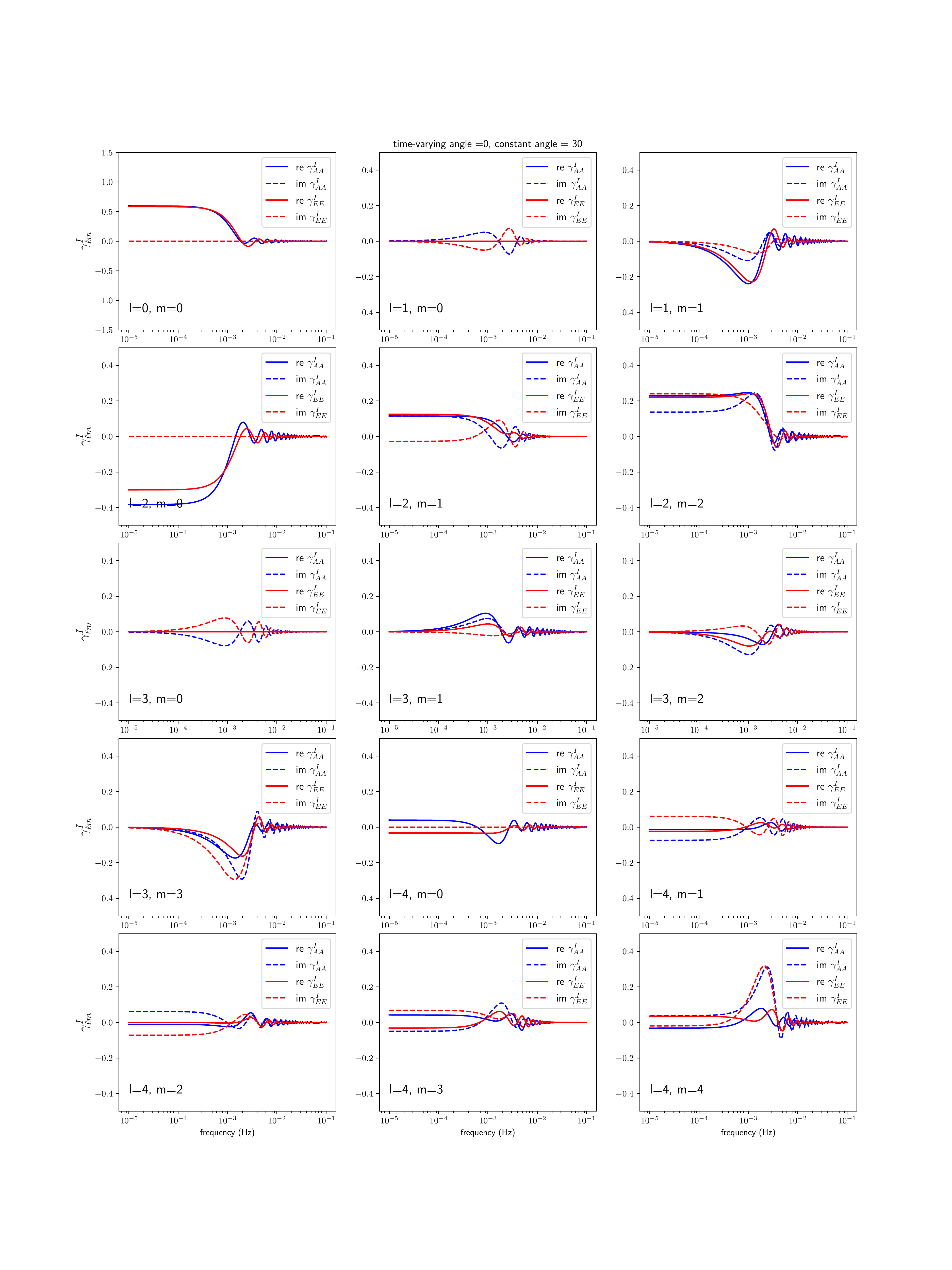}
    \caption[Coordinate]{Same as in Fig.~\ref{fig:ORFI0AAEE} except for $\gamma_{\ell m}^I$ when $\phi(t)=0^\circ$ and the constant angle $\phi_0=30^\circ$.}
\label{fig:ORFI30AAEE}
\index{figures}
\end{figure}

\begin{figure}[ht!]
	\centering
	\includegraphics[width=1.0\textwidth]{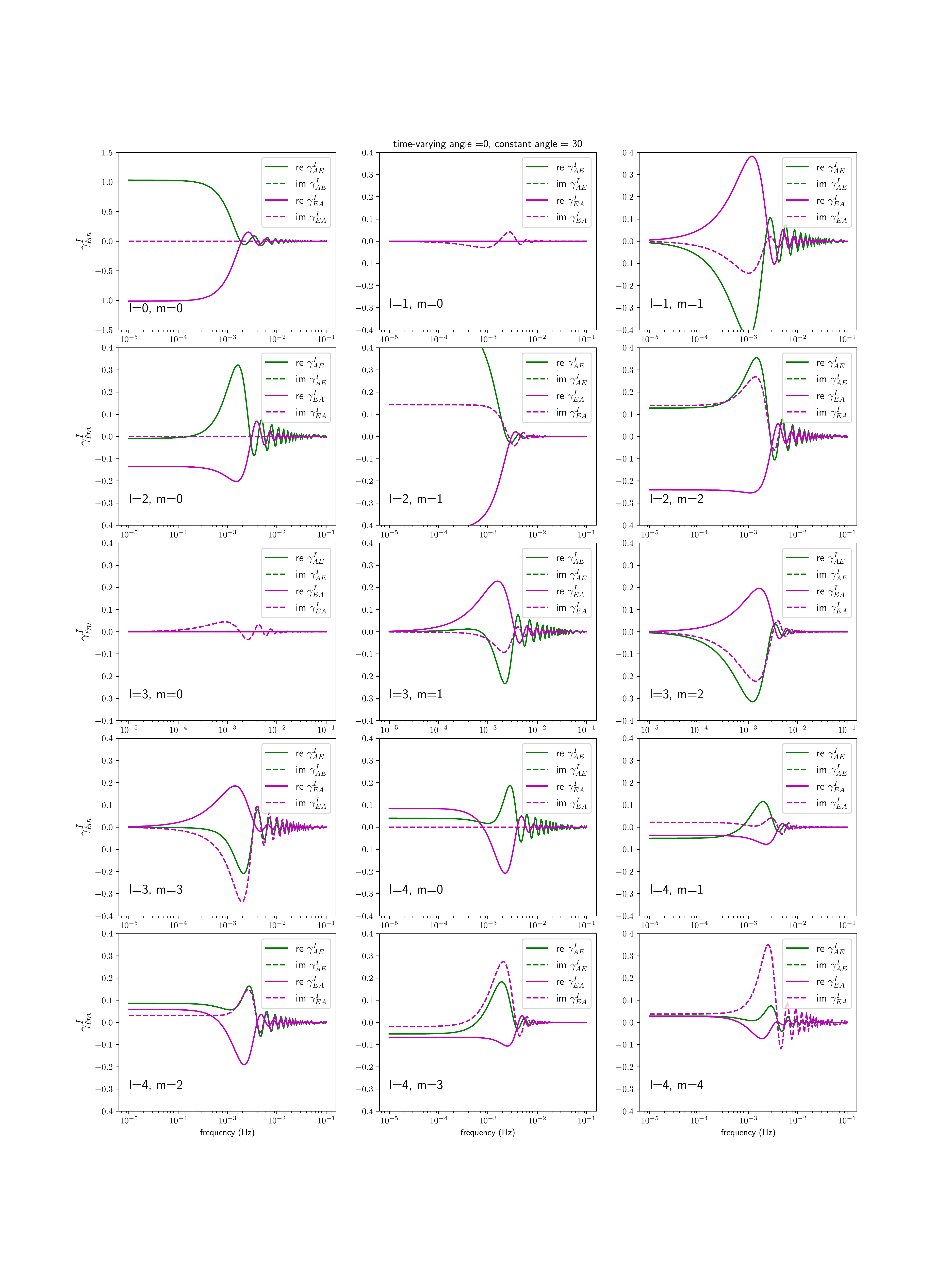}
	\caption[Coordinate]{Same as in Fig.~\ref{fig:ORFI0AEEA} except for $\gamma_{\ell m}^I$ when $\phi(t)=0^\circ$ and the constant angle $\phi_0=30^\circ$.}
	\label{fig:ORFI30AEEA}
	\index{figures}
\end{figure}

\begin{figure}[ht!]
\centering
\includegraphics[width=1.0\textwidth]{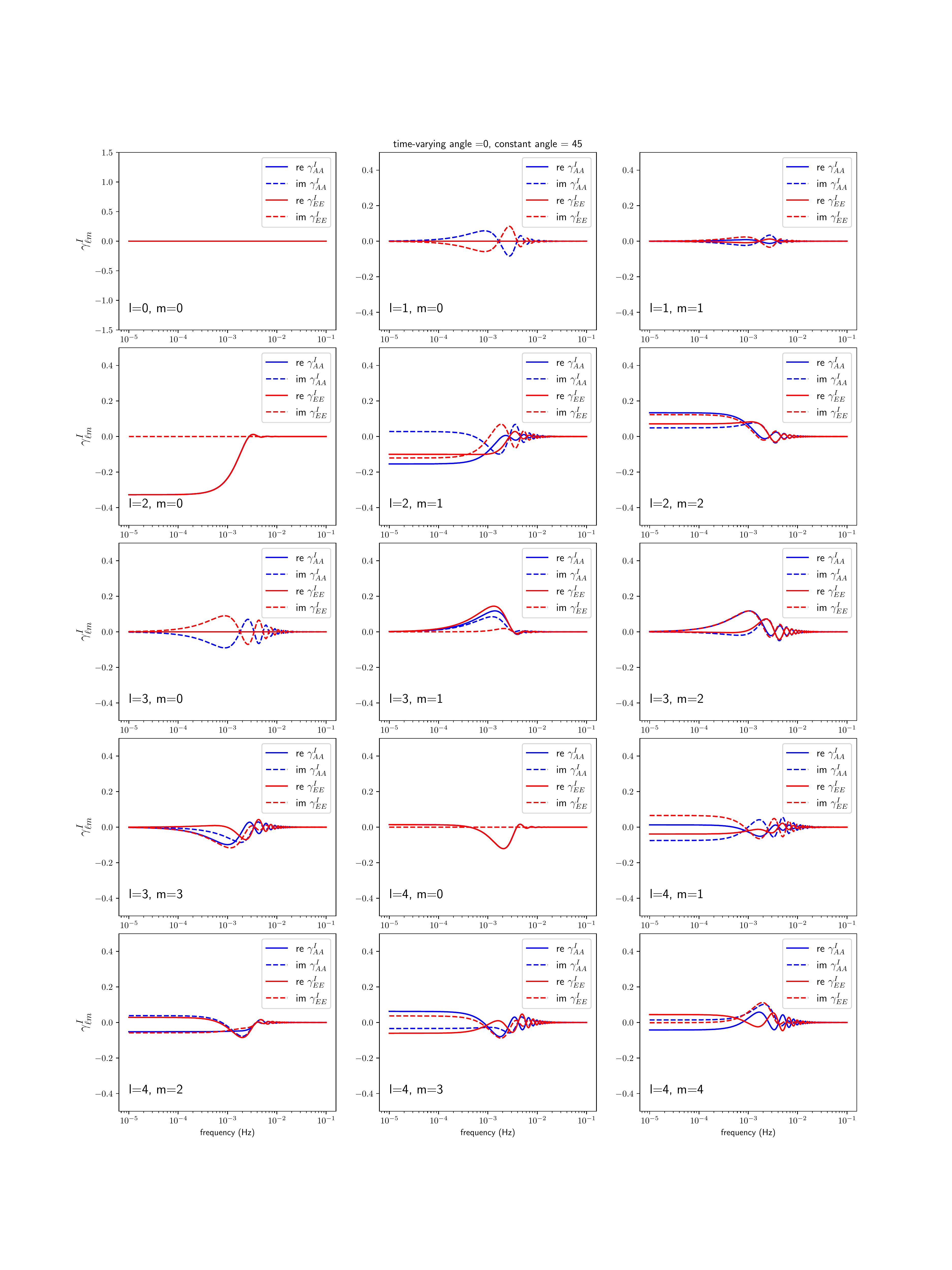}
    \caption[Coordinate]{Same as in Fig.~\ref{fig:ORFI0AAEE} except for $\gamma_{\ell m}^I$ when $\phi(t)=0^\circ$ and the constant angle $\phi_0=45^\circ$.}
\label{fig:ORFI45AAEE}
\index{figures}
\end{figure}

\begin{figure}[ht!]
	\centering
	\includegraphics[width=1.0\textwidth]{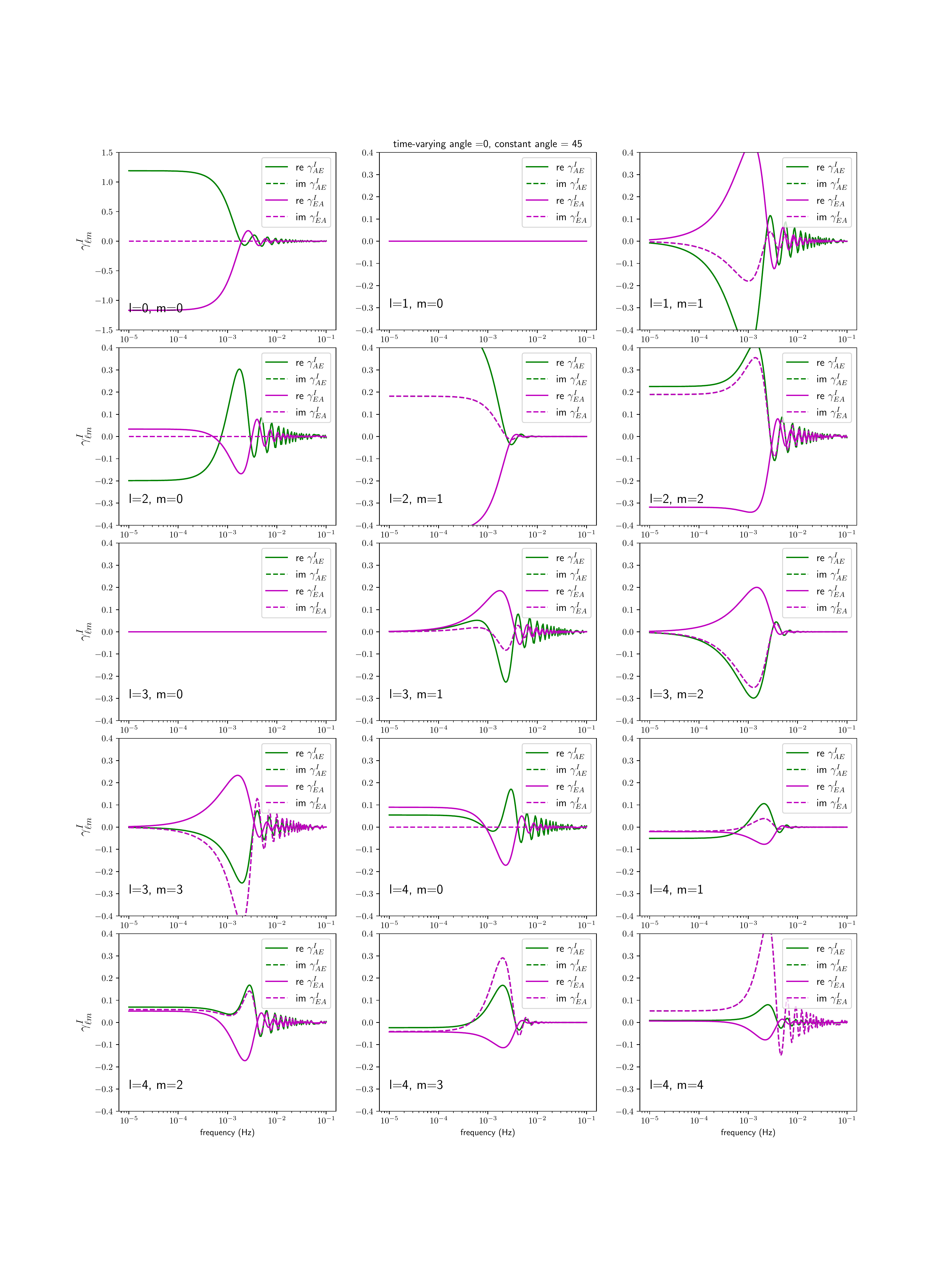}
	\caption[Coordinate]{Same as in Fig.~\ref{fig:ORFI0AEEA} except for $\gamma_{\ell m}^I$ when $\phi(t)=0^\circ$ and the constant angle $\phi_0=45^\circ$.}
	\label{fig:ORFI45AEEA}
	\index{figures}
\end{figure}

\begin{figure}[ht!]
\centering
\includegraphics[width=1.0\textwidth]{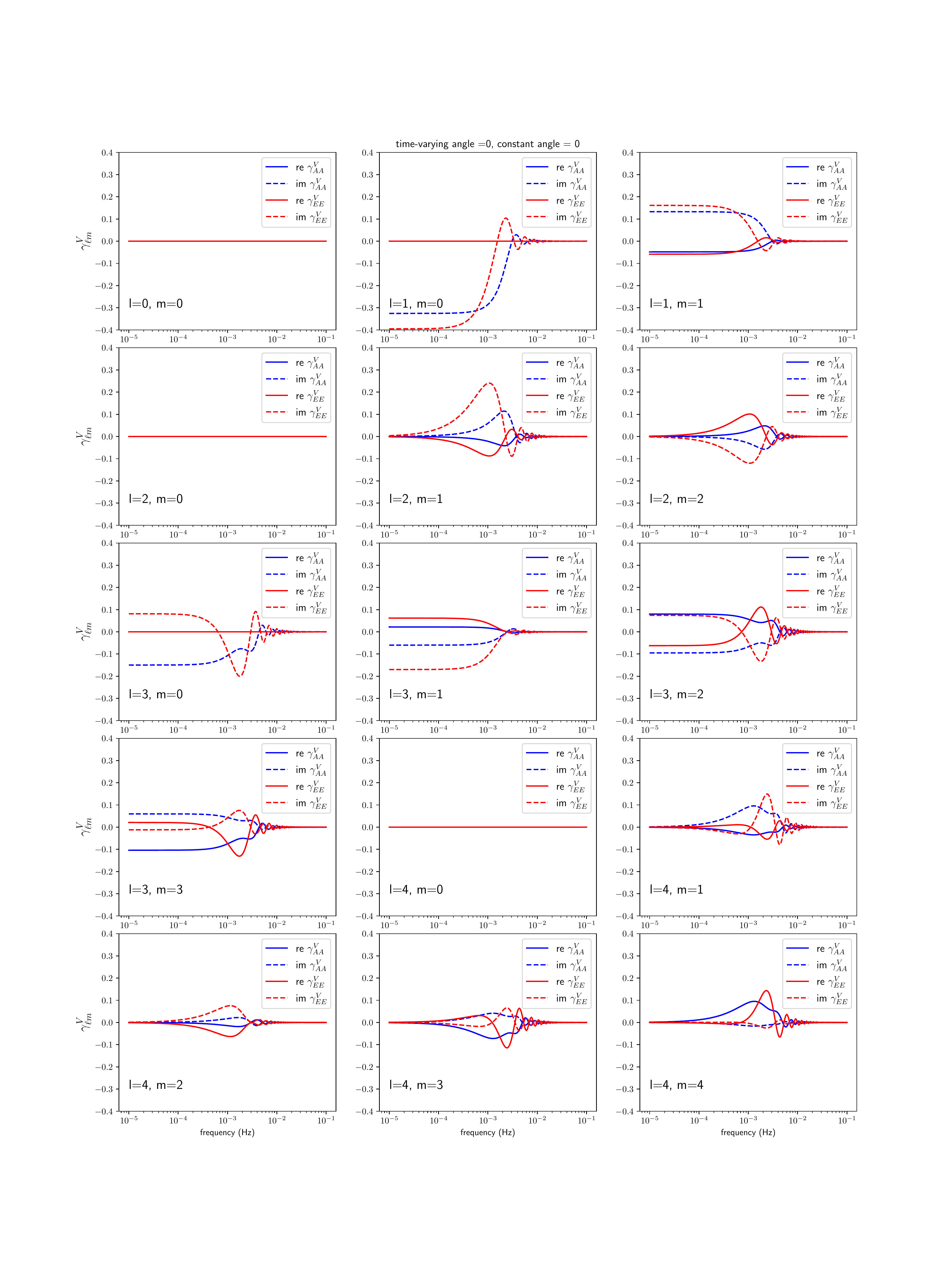}
    \caption[Coordinate]{Real and imaginary parts of the multipole moments of the circular-polarization $A$-$A$ and $E$-$E$ overlap reduction function $\gamma_{\ell m}^V$ for the LISA-Taiji detector pair when $\phi(t)=0^\circ$ and the constant angle 
    $\phi_0=0^\circ$. Plots of $\ell=0,1,2,3,4$ and $m\ge 0$ are shown. The $m<0$ multipoles can be obtained by using the conjugate relation in Eq.~(\ref{IVconjugate}).}
\label{fig:ORFV0AAEE}
\index{figures}
\end{figure}

\begin{figure}[ht!]
	\centering
\includegraphics[width=1.0\textwidth]{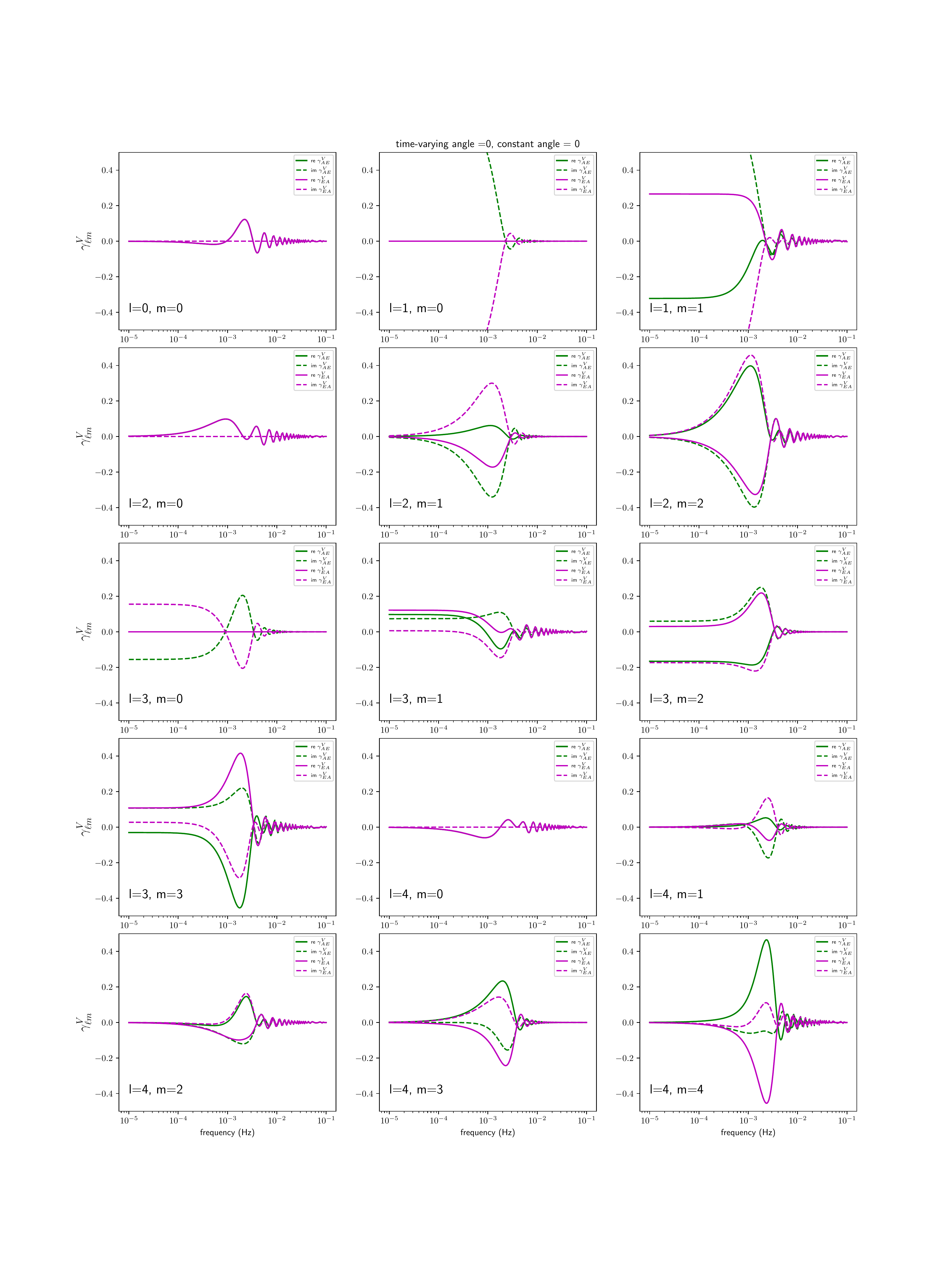}
\caption[Coordinate]{Real and imaginary parts of the multipole moments of the circular-polarization $A$-$E$ and $E$-$A$ overlap reduction function $\gamma_{\ell m}^V$ for the LISA-Taiji detector pair when $\phi(t)=0^\circ$ and the constant angle 
	$\phi_0=0^\circ$. Plots of $\ell=0,1,2,3,4$ and $m\ge 0$ are shown. The $m<0$ multipoles can be obtained by using the conjugate relation in Eq.~(\ref{IVconjugate}).}
\label{fig:ORFV0AEEA}
\index{figures}
\end{figure}

\begin{figure}[ht!]
\centering
\includegraphics[width=1.0\textwidth]{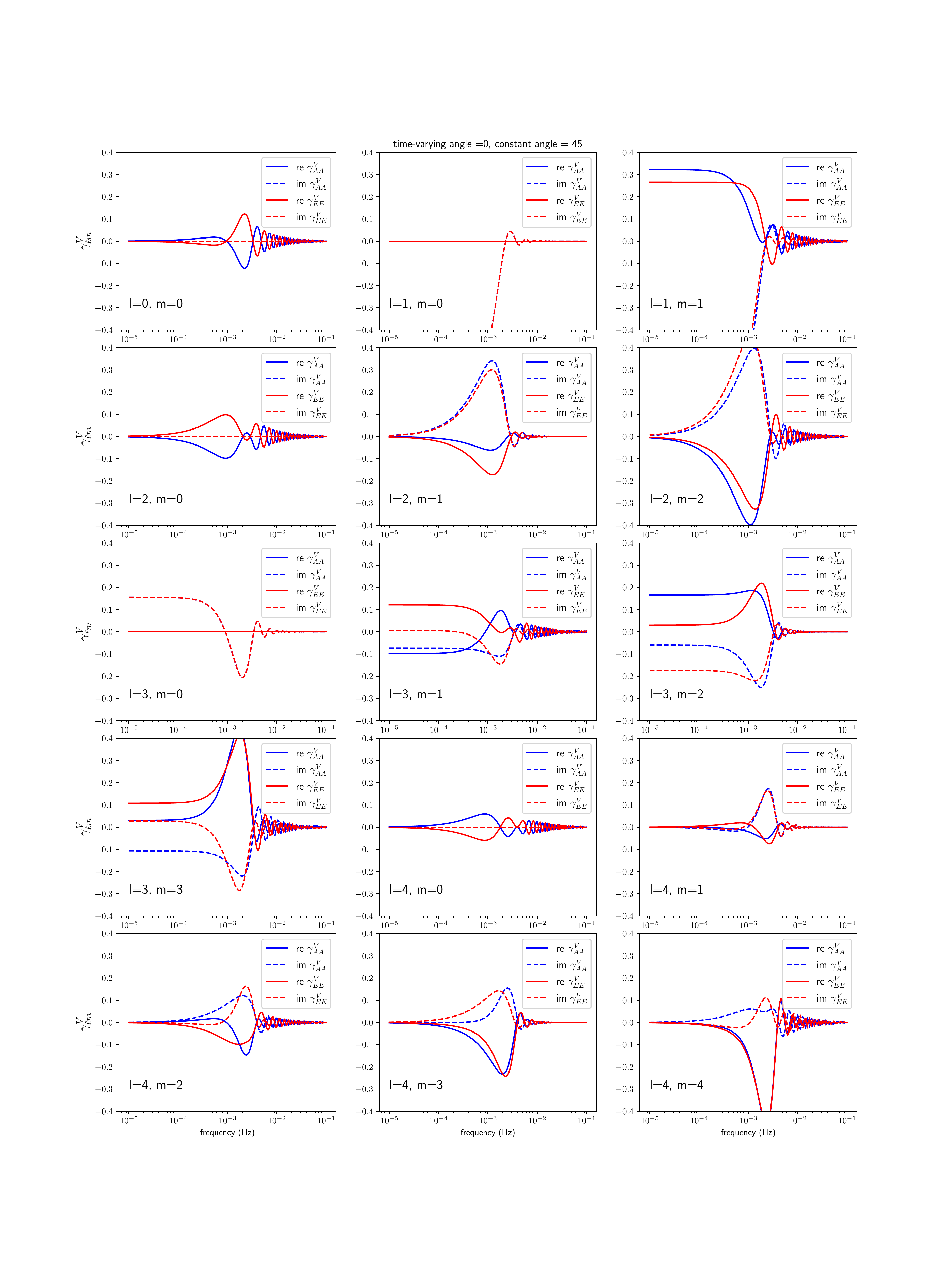}
    \caption[Coordinate]{Same as in Fig.~\ref{fig:ORFV0AAEE} except for $\gamma_{\ell m}^V$ when $\phi(t)=0^\circ$ and the constant angle $\phi_0=45^\circ$.}
\label{fig:ORFV45AAEE}
\index{figures}
\end{figure}

\begin{figure}[ht!]
	\centering
	\includegraphics[width=1.0\textwidth]{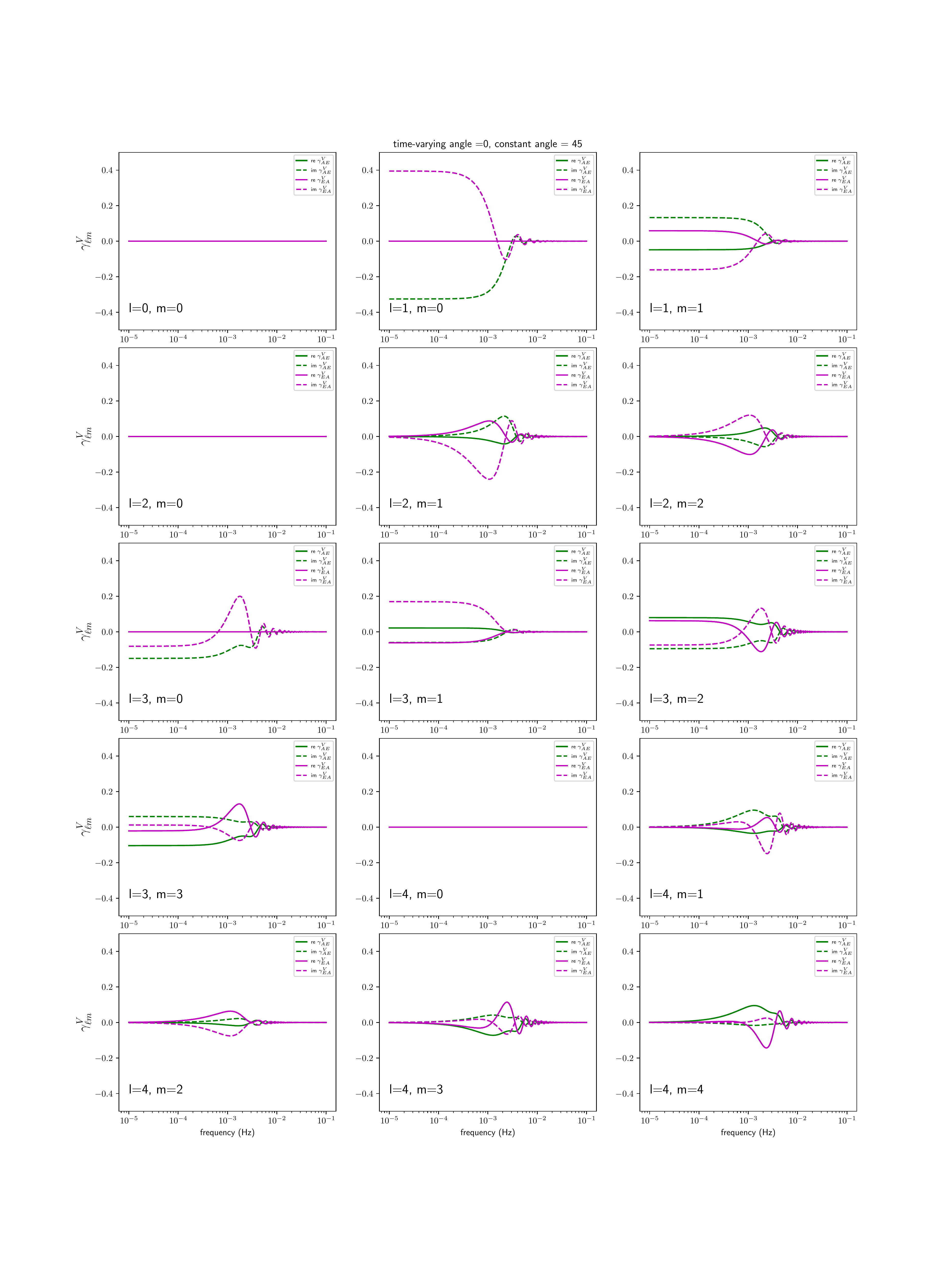}
	\caption[Coordinate]{Same as in Fig.~\ref{fig:ORFV0AEEA} except for $\gamma_{\ell m}^V$ when $\phi(t)=0^\circ$ and the constant angle $\phi_0=45^\circ$.}
	\label{fig:ORFV45AEEA}
	\index{figures}
\end{figure}

\begin{figure}[ht!]
\centering
\includegraphics[width=1.0\textwidth]{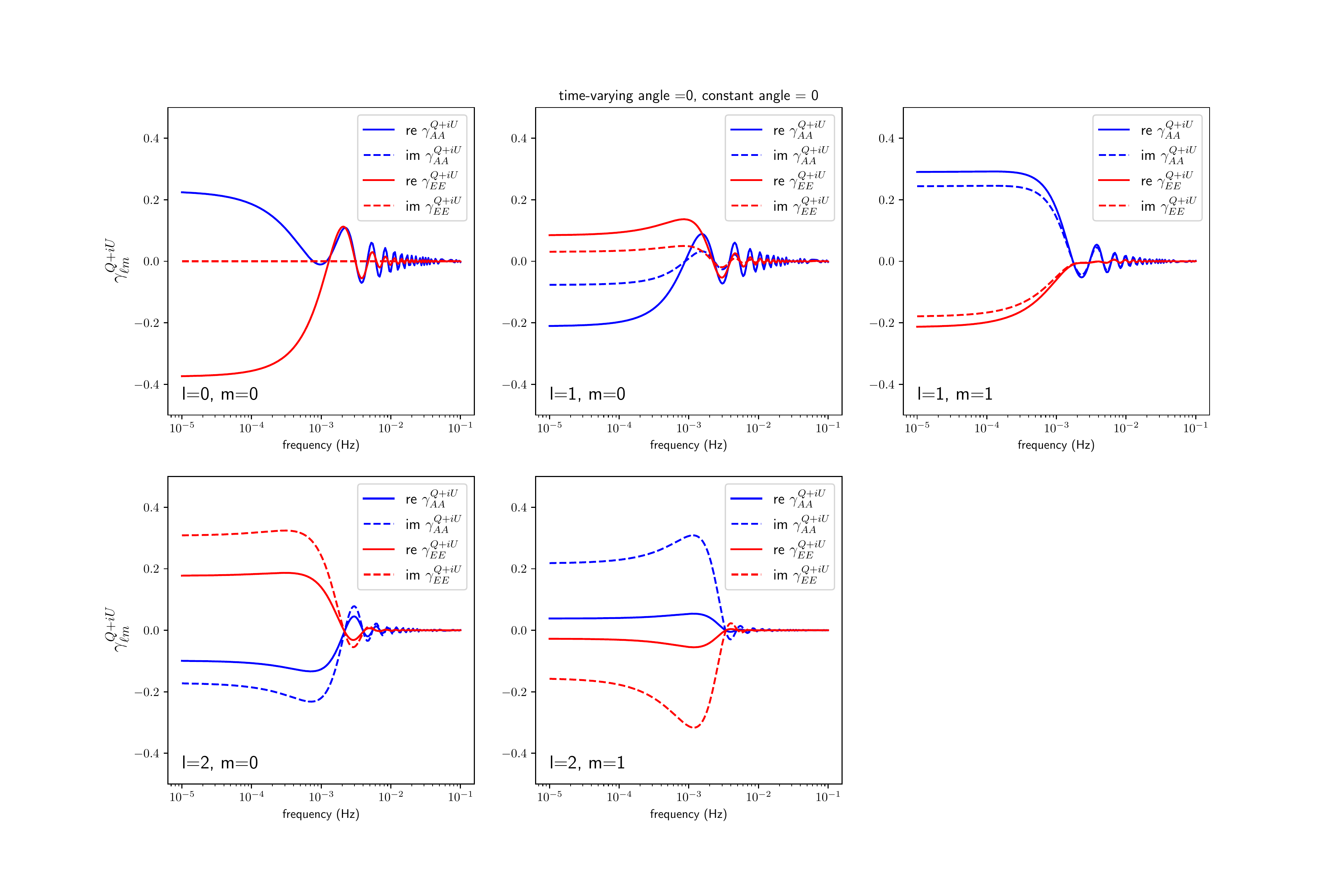}
\includegraphics[width=1.0\textwidth]{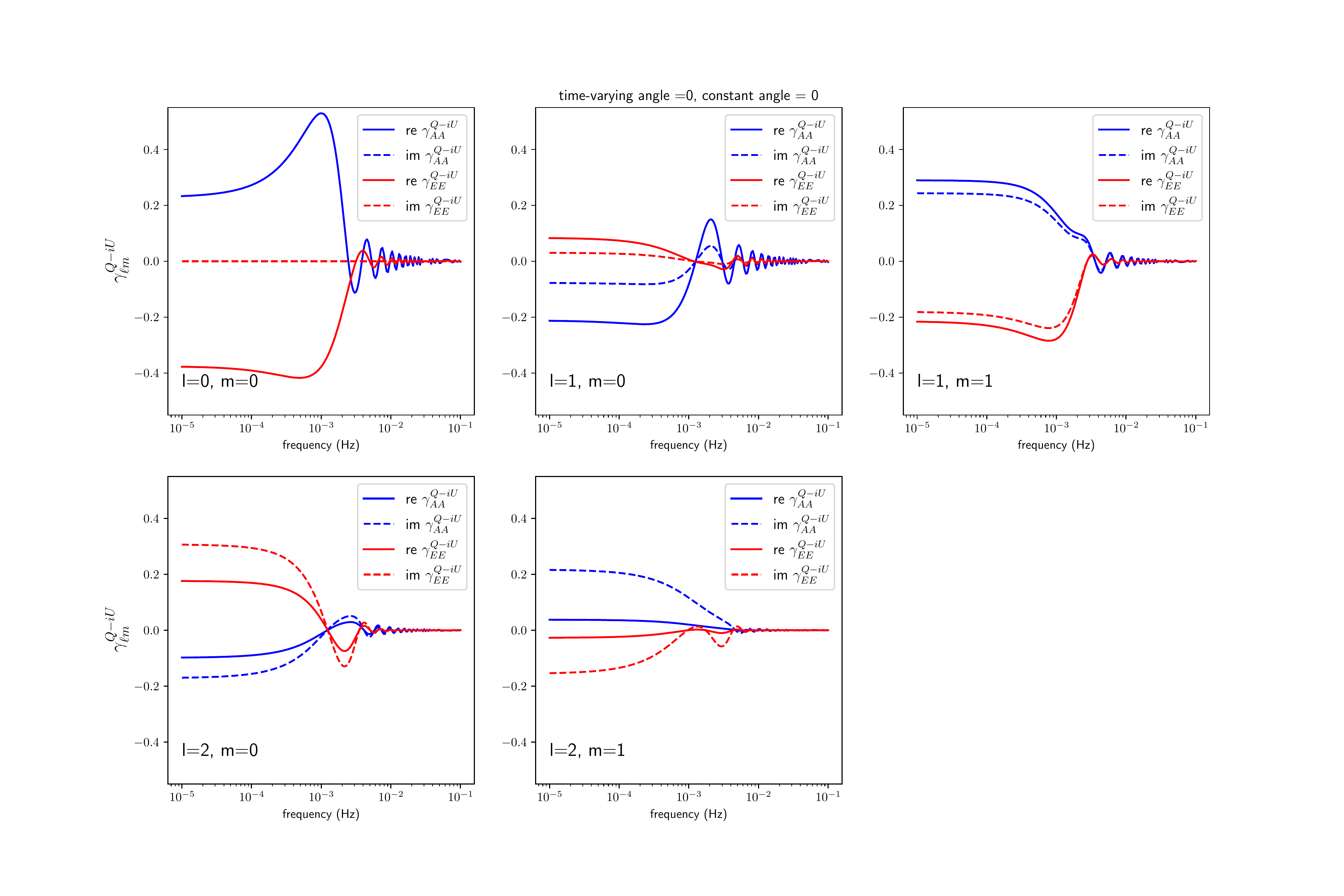}
    \caption[Coordinate]{Real and imaginary parts of the multipole moments of the linear-polarization
    $A$-$A$ and $E$-$E$  overlap reduction function $\gamma_{\ell m}^{Q\pm iU}$ for the LISA-Taiji detector pair when $\phi(t)=0^\circ$ and the constant angle $\phi_0=0^\circ$. Plots of $\ell=4$ and $m=0,1,2,3,4$ are shown. The $m<0$ multipoles can be obtained by using the conjugate relations in Eq.~(\ref{QUconjugate}).}
\label{fig:ORFQU0AAEE}
\index{figures}
\end{figure}

\begin{figure}[ht!]
	\centering
	\includegraphics[width=1.0\textwidth]{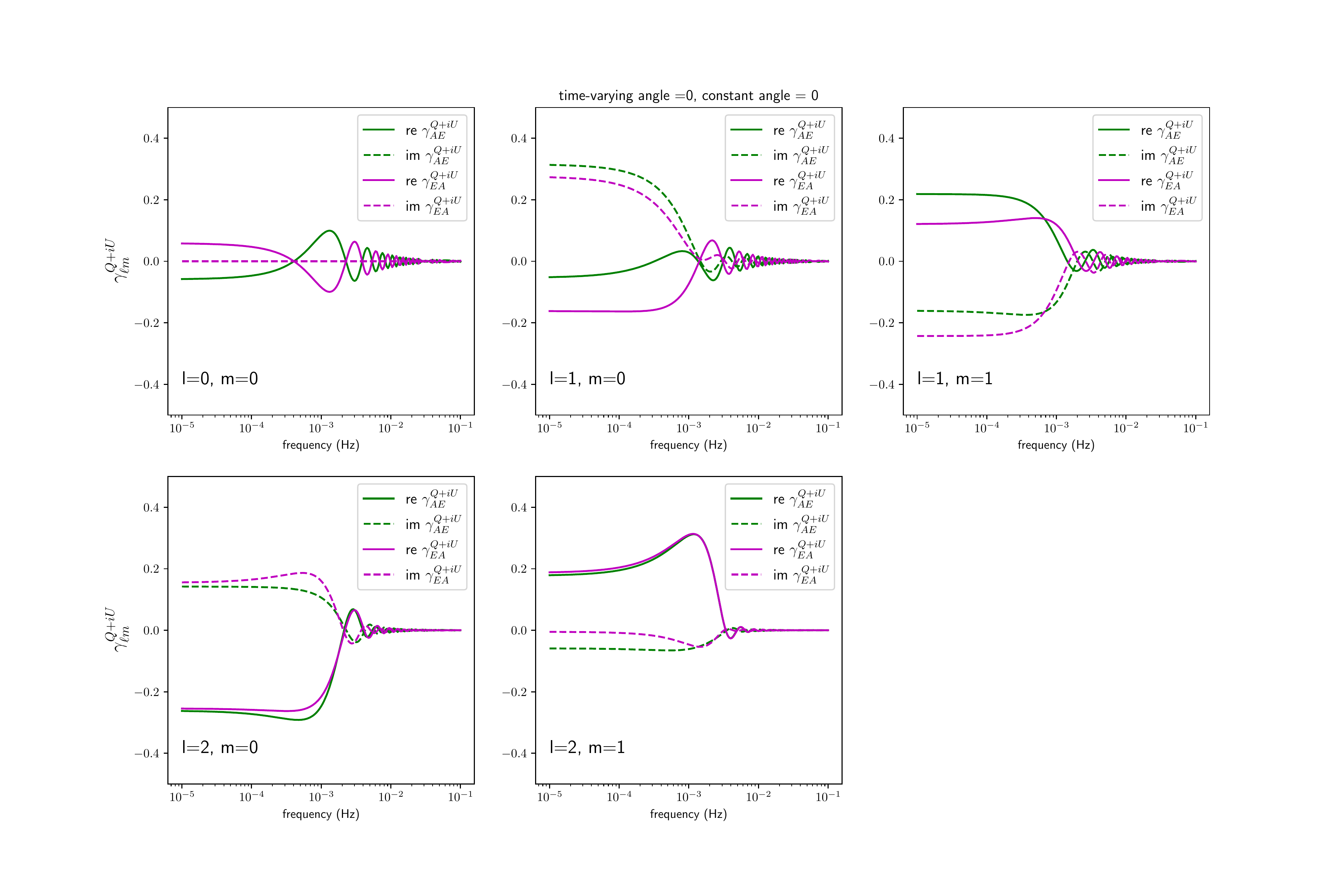}
	\includegraphics[width=1.0\textwidth]{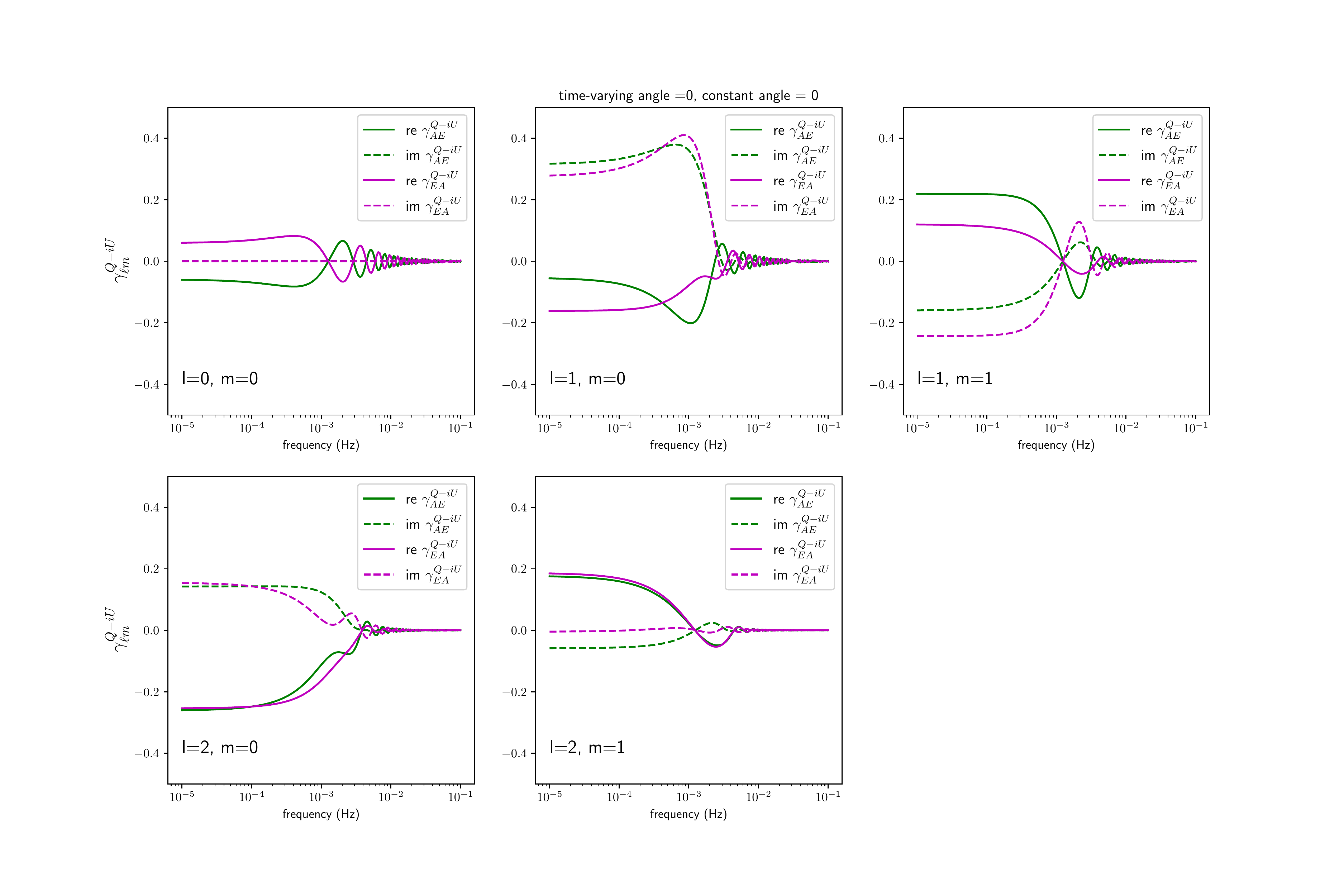}
	\caption[Coordinate]{Real and imaginary parts of the multipole moments of the linear-polarization
		$A$-$E$ and $E$-$A$  overlap reduction function $\gamma_{\ell m}^{Q\pm iU}$ for the LISA-Taiji detector pair when $\phi(t)=0^\circ$ and the constant angle $\phi_0=0^\circ$. Plots of $\ell=4$ and $m=0,1,2,3,4$ are shown. The $m<0$ multipoles can be obtained by using the conjugate relations in Eq.~(\ref{QUconjugate}).}
	\label{fig:ORFQU0AEEA}
	\index{figures}
\end{figure}

To compute the ORF multipoles in Eqs.~(\ref{eq:gammaIV_lm}) and~(\ref{eq:gammaQU_lm}), one need the Wigner-D matrices $D^{\ell}_{m' m}(-\alpha,-\theta_a,-\phi_a)$, where $(\theta_a,\phi_a)$ is the spherical coordinates of the LISA and $\alpha$ is the angle between its meridian and the great circle connecting LISA-Taiji. 
From Fig.~\ref{fig:LisaTaiji}, $\theta_a=120^\circ$ and $\alpha=79.7^\circ$. 
Without loss of generality, we choose $\phi_a=\phi(t)$, because the orbital and spinning periods are the same. 
Overall, the ORF multipoles are dependent on a single time-varying angle $\phi(t)$.
We have made some plots of the ORF multipoles in Figs.~\ref{fig:ORFI0AAEE}-\ref{fig:ORFI45AEEA} for 
$\gamma_{\ell m}^I$, Figs.~\ref{fig:ORFV0AAEE}-\ref{fig:ORFV45AEEA} for $\gamma_{\ell m}^V$, and Figs.~\ref{fig:ORFQU0AAEE}-\ref{fig:ORFQU0AEEA} for $\gamma_{\ell m}^{Q\pm iU}$, by selecting some representative values for $\phi(t)$ and $\phi_0$. 

For an isotropic unpolarized SGWB, it is expected that it carries a dipole anisotropy due to the Doppler motion 
of the detector with respect to the background.
Figure.~\ref{fig:ORFI0AEEA} shows that the $E$-$A$ correlation can select this dipole component. 
Furthermore, we can use the $A$-$E$ correlation to detect quadrupole or higher-multipole anisotropy, if there is any.
Similarly, Fig.~\ref{fig:ORFI45AAEE} shows that either $A$-$A$ or $E$-$E$ correlation can select the dipole component. 
Taking the sum of $A$-$A$ and $E$-$E$ correlations can detect quadrupole or higher-multipole anisotropy.

For an isotropic circularly polarized SGWB,
in Figs.~\ref{fig:ORFI45AAEE} and \ref{fig:ORFV45AAEE}, taking the sum of $A$-$A$ and $E$-$E$ correlations can single out the circular-polarization dipole component. 

\subsection{Time-independent antenna pattern functions}
\label{newORF}

For the ORFs in the LISA-Taiji network, we have identified a single time-varying angle $\phi(t)$ for the orbiting spacecrafts and a phase angle difference $\phi_0$ between the two triangular detectors. 
However, it would be more convenient to use time-independent antenna pattern functions in doing the correlation.
This can be done by introducing a phase rotation to the detector signals, similar to a strategy adopted in full-sky observation of the CMB anisotropy and polarization correlation~\cite{ngliu}. As such, the LISA detector tensors become
\begin{align}
    \begin{pmatrix}
        d_{A'} \\
        d_{E'}    
     \end{pmatrix} =
     \begin{pmatrix}
        \cos 2\phi(t) && \sin 2\phi(t) \\
         -\sin 2\phi(t)&&  \cos 2\phi(t)
    \end{pmatrix}
    \begin{pmatrix}
       d_{A} \\
        d_{E}    
    \end{pmatrix} \,,
\end{align}
and the Taiji's are given by
\begin{align}
    \begin{pmatrix}
        d_{A'} \\
        d_{E'}    
     \end{pmatrix} =
     \begin{pmatrix}
        \cos 2[\phi(t)-\phi_0]&& \sin 2[\phi(t)-\phi_0] \\
         -\sin 2[\phi(t)-\phi_0]&&  \cos 2[\phi(t)-\phi_0]
    \end{pmatrix}
    \begin{pmatrix}
       d_{A} \\
        d_{E}    
    \end{pmatrix} \,.
\end{align}
This rotation makes the X arms of the LISA's and Taiji's $A$ channels be aligned 
with the great circle connecting LISA-Taiji. Hence, the resulting ORFs for an isotropic SGWB 
are given by the special case in Eqs.~(\ref{eq:gammaI_iso2}) and~(\ref{eq:gammaV_iso2}) 
when $\phi(t)=\phi_0=0$:
\begin{align}
    \gamma_{00,A'A'}^{I}(f) 
    =&
    C_1(f)  - C_2(f) \,, \nonumber \\
  \gamma_{00,E'E'}^{I}(f) 
    =&
    C_1(f)   + C_2(f) \,, \nonumber \\
 \gamma_{00,A'E'}^{V}(f) =&  \gamma_{00,E'A'}^{V}(f) = C_3(f) \,, 
\end{align}
which correspond to the so-called virtually aligned channels that have been used to separate the $I_{00}$ and $V_{00}$ components~\cite{seto2020}. Furthermore, the ORF multipoles for $A'$ and $E'$ correlations are given 
by those in Sec.~\ref{ORFmultipole} in the case with $\phi(t)=\phi_0=0$.

\section{Correlation signal and frequency filter}

The LISA-Taiji network has an uniform sky coverage that is a circular ring about the ecliptic pole resulted from the Earth's revolution. 
The correlation output is periodic in the azimuthal angle $\phi_a=\phi(t)$ with a period of $1\,{\rm yr}$, 
given by Eqs.~(\ref{eq:xilm}), (\ref{eq:gammaIV_lm}),  (\ref{eq:gammaQU_lm}), and (\ref{eq:DtoSWSH}) as
\begin{align}
   \xi_{ab}(f,t)
   &=
    \sum_{S}
   \sum_{\ell m}  S_{\ell m}(f) \gamma_{\ell m}^{S}(f) \,e^{im\phi(t)}\,,
\end{align}
where $\gamma_{\ell m}^{S}$ are the ORF multipoles for $A'$ and $E'$ correlations in Sec.~\ref{newORF}.
We can use a filter function of frequency $Q(f)$ to optimize the signal as~\cite{allen1997}
\begin{equation}
S(t)=\int_{-\infty}^{\infty} \d f \, \xi_{ab}(f,t) \,Q(f)\,.
\end{equation}
Then, the Fourier mode of $S(t)$ is given by
\begin{align}
S_m=&\frac{1}{2\pi}\int_0^{2\pi} \d \phi(t) \,e^{-im\phi(t)}S(t) \\ \nonumber
       =& \int_{-\infty}^{\infty} \d f \, Q(f)\, \sum_{S}
   \sum_{\ell=|m|}^\infty  S_{\ell m}(f) \gamma_{\ell m}^{S}(f)\,.
\end{align}
It was shown that to maximize the signal to noise ratio for $S_m$, 
the optimal choice of the filter function,
here generalized to including SGWB polarization anisotropies,
is given by~\cite{allen1997}
\begin{equation}
Q(f) =\frac{1}{N_a(f) N_b(f)}\sum_{S}
   \sum_{\ell=|m|}^\infty  S_{\ell m}^*(f) \gamma_{\ell m}^{S*}(f)\,,
\end{equation}
where $N_a(f)$ and $N_b(f)$ are the LISA's and Taiji's noise power spectra, respectively.

The situation in the ET-CE network is similar. The sky coverage is a circular ring about the celestial pole
and the period is $1\,{\rm day}$ due to the Earth's rotation. For the ET-CE pair, $\gamma_{\ell m}^{S}$ are the ORF multipoles given in Sec.~\ref{ETCEpair}, with $\phi_a=0$.

\bw


\ew

\section{Conclusion}
\label{conclusion}

We have discussed using future gravitational-wave interferometers to form detector pairs for measuring the properties of the stochastic gravitational-wave background, such as the Einstein Telescope-Cosmic Explorer detector pair and the LISA-Taiji network. We have used a fast algorithm developed in Ref.~\cite{chu21} to obtain the overlap reduction functions for an anisotropic polarized stochastic gravitational-wave background. The overlap reduction functions of the Einstein Telescope-Cosmic Explorer pair are similar to those of the pair formed by LIGO-Hanford and LIGO-Livingston as obtained in Ref.~\cite{chu21}. For the overlap reduction functions in the LISA-Taiji network, we have identified a time-varying angle $\phi(t)$ and a phase angle difference $\phi_0$ between the two triangular detectors. We emphasize that since the configurations of these detectors have not been finalized, the results in the present work are very useful to the detector and mission design targeting at stochastic gravitational-wave background measurements.

\begin{acknowledgments}
This work was supported in part by the Ministry of Science and Technology (MOST) of Taiwan, Republic of
China, under Grants No. MOST 110-2112-M-032-007 (G.C.L.), No. MOST 110-2123-M-007-002 (G.C.L.), and No. MOST 111-2112-M-001-065 (K.W.N.).
\end{acknowledgments}

\appendix
\bw
\section{Spin-Weighted Spherical Harmonics}
The explicit form of the spin-weighted spherical harmonics that we use is
\begin{align}
{}_{s}Y_{\ell m}(\theta,\phi) = 
    (-1)^{m}
    e^{im\phi}
    \sqrt{\frac{(2\ell+1)}{(4\pi)}\frac{(\ell+m)!(\ell-m)!}{(\ell+s)!(\ell-s)!}}
    \sin^{2\ell}\!\left(\frac{\theta}{2}\right)
    \sum_{r}
    \binom{\ell-s}{r} \binom{\ell+s}{r+s-m}
    (-1)^{\ell-r-s}
    \cot^{2r+s-m}\!\left(\frac{\theta}{2}\right) \,.
\end{align}
\ew
When $s=0$, it reduces to the ordinary spherical harmonics,
\be
Y_{\ell m}(\hat{n})=\sqrt{\frac{(2\ell+1)}{(4\pi)}\frac{(\ell-m)!}{(\ell+m)!}}P_{\ell m}(\cos \theta) e^{i m \phi} \,.
\ee

Spin-weighted spherical harmonics satisfy the orthogonal relation,
\be
    \int_{S^2} \d{\hat{n}}\; {}_{s}Y^*_{\ell m}(\hat{n}){}_{s}Y_{\ell' m'}(\hat{n})
    = \delta_{\ell \ell'} \delta_{m m'} \,,
\ee
and the completeness relation,
\begin{align}
    \sum_{\ell m} {}_{s}Y^*_{\ell m}(\hat{n}){}_{s}Y_{\ell m}(\hat{n}')
    =& \delta(\hat{n}-\hat{n}') \nonumber \\
    =& \delta(\phi-\phi')\delta(\cos\theta-\cos\theta')\,.
\end{align}
Its complex conjugate is
\be
{}_{s}Y^*_{\ell m}(\hat{n}) =  (-1)^{s+m} {}_{-s}Y_{\ell -m}(\hat{n}) \,,
\label{Yconjugate}
\ee
and its parity is given by
\be
{}_{s}Y_{\ell m}(-\hat{n}) \equiv {}_{s}Y_{\ell m}(\pi-\theta,\phi+\pi)=(-1)^{\ell} {}_{-s}Y_{\ell m}(\hat{n}) \,.
\label{Yparity}
\ee

\bw
\section{Antenna Pattern Functions}
\label{sec:DE}
In most literature, the inner product between the detector tensor and the polarization basis tensor is referred as the antenna pattern function. 

\subsection{$\mathbb{DE}^I$}
\label{sec:DEI}
For the Stokes-I parts, the only non-vanishing $\mathbb{D}_0(\sigma_1,\sigma_2,\beta) \cdot \mathbb{E}^{I}_{\ell m}$ are 15 components with $\ell=0,2,4$, which satisfy $\mathbb{DE}_{\ell -m}=(-1)^m \mathbb{DE}^*_{\ell m}$.
\begin{align*}
    \mathbb{DE}^I_{00}
    &=
    \frac{4}{5} \sqrt{\pi } \left(\cos ^4\left(\frac{\beta }{2}\right) \cos (2 \sigma_1- 2 \sigma_2 )+\sin ^4\left(\frac{\beta }{2}\right) \cos (2 \sigma_1+ 2 \sigma_2 )\right)
    \\
    \mathbb{DE}^I_{20}
    &=
    \frac{8}{7} \sqrt{\frac{\pi }{5}} \left(\cos ^4\left(\frac{\beta }{2}\right) \cos (2 \sigma_1- 2 \sigma_2 )+\sin ^4\left(\frac{\beta }{2}\right) \cos (2 \sigma_1+ 2 \sigma_2 )\right)
    \\
    \mathbb{DE}^I_{21}
    &=
    -\frac{2}{7} \sqrt{\frac{6 \pi }{5}} e^{-2 i \sigma_1} \sin (\beta ) (\cos (\beta ) \cos (2 \sigma_2)+i \sin (2 \sigma_2))
    \\
    \mathbb{DE}^I_{22}
    &=
    \frac{2}{7} \sqrt{\frac{6 \pi }{5}} e^{-2 i \sigma_1} \sin ^2(\beta ) \cos (2 \sigma_2)
    \\
    \mathbb{DE}^I_{40}
    &=
    \frac{2}{105} \sqrt{\pi } \left(\cos ^4\left(\frac{\beta }{2}\right) \cos (2 \sigma_1 - 2 \sigma_2)+\sin ^4\left(\frac{\beta }{2}\right) \cos (2 \sigma_1 + 2 \sigma_2 )\right)
    \\
    \mathbb{DE}^I_{41}
    &=
    -\frac{1}{21} \sqrt{\frac{\pi }{5}} e^{-2 i \sigma_1} \sin (\beta ) (\cos (\beta ) \cos (2 \sigma_2)+i \sin (2 \sigma_2))
    \\
    \mathbb{DE}^I_{42}
    &=
    \frac{1}{7} \sqrt{\frac{\pi }{10}} e^{-2 i \sigma_1} \sin ^2(\beta ) \cos (2 \sigma_2)
    \\
    \mathbb{DE}^I_{43}
    &=
    \frac{1}{3} \sqrt{\frac{\pi }{35}} e^{-2 i \sigma_1} \sin (\beta ) (\cos (\beta ) \cos (2 \sigma_2)-i \sin (2 \sigma_2))
    \\
    \mathbb{DE}^I_{44}
    &=
    \frac{1}{6} \sqrt{\frac{\pi }{70}} e^{-2 i \sigma_1} ((\cos (2 \beta )+3) \cos (2 \sigma_2)-4 i \cos (\beta ) \sin (2 \sigma_2))
\end{align*}

\subsection{$\mathbb{DE}^V$}
\label{sec:DEV}
For the Stokes-V parts that correspond to the circular polarized signal, the only non-vanishing $\mathbb{D}_0(\sigma_1,\sigma_2,\beta) \cdot \mathbb{E}^{V}_{\ell m}$ are 10 components with $\ell=1,3$, which satisfy $\mathbb{DE}_{\ell -m}=(-1)^{m+1} \mathbb{DE}^*_{\ell m}$.

\begin{align*}
    \mathbb{DE}^V_{10}
    &=
    \frac{8}{5} i \sqrt{\frac{\pi }{3}} \left(\cos ^4\left(\frac{\beta }{2}\right) \sin (2 \sigma_1 - 2 \sigma_2 )+\sin ^4\left(\frac{\beta }{2}\right) \sin (2 \sigma_1 + 2\sigma_2 )\right)
    \\
    \mathbb{DE}^V_{11}
    &=
    \frac{2}{5} \sqrt{\frac{2 \pi }{3}} e^{-2 i \sigma_1} \sin (\beta ) (\cos (\beta ) \cos (2 \sigma_2)+i \sin (2 \sigma_2))
    \\
    \mathbb{DE}^V_{30}
    &=
    \frac{2}{5} i \sqrt{\frac{\pi }{7}} \left(\cos ^4\left(\frac{\beta }{2}\right) \sin (2 \sigma_1 - 2 \sigma_2 )+\sin ^4\left(\frac{\beta }{2}\right) \sin (2 \sigma_1 + 2 \sigma_2 )\right)
    \\
    \mathbb{DE}^V_{31}
    &=
    \frac{1}{5} \sqrt{\frac{3 \pi }{7}} e^{-2 i \sigma_1} \sin (\beta ) (\cos (\beta ) \cos (2 \sigma_2)+i \sin (2 \sigma_2))
    \\
    \mathbb{DE}^V_{32}
    &=
    -\sqrt{\frac{3 \pi }{70}} e^{-2 i \sigma_1} \sin ^2(\beta ) \cos (2 \sigma_2)
    \\
    \mathbb{DE}^V_{33}
    &=
    -\sqrt{\frac{\pi }{35}} e^{-2 i \sigma_1} \sin (\beta ) (\cos (\beta ) \cos (2 \sigma_2)-i \sin (2 \sigma_2))
\end{align*}

\subsection{$\mathbb{DE}^{Q\pm iU}$}
For the linear polarized signal, the only non-vanishing $\mathbb{D}_0(\sigma_1,\sigma_2,\beta) \cdot \mathbb{E}^{Q\pm i U}_{\ell m}$ are 9 components with $\ell=4$, which satisfy $\mathbb{DE}_{\ell -m}=(-1)^m \mathbb{DE}^*_{\ell m}$.
\begin{align*}
    \mathbb{DE}^{Q\pm iU}_{40}
    &=
    \frac{1}{3} \sqrt{\frac{2 \pi }{35}} \left(\cos ^4\left(\frac{\beta }{2}\right) \cos (2 \sigma_1- 2 \sigma_2 )+\sin ^4\left(\frac{\beta }{2}\right) \cos (2 \sigma_1+ 2 \sigma_2)\right)
    \\
    \mathbb{DE}^{Q\pm iU}_{41}
    &=
    -\frac{1}{3} \sqrt{\frac{\pi }{14}} e^{-2 i \sigma_1} \sin (\beta ) (\cos (\beta ) \cos (2 \sigma_2)+i \sin (2 \sigma_2))
    \\
    \mathbb{DE}^{Q\pm iU}_{42}
    &=
    \frac{1}{2} \sqrt{\frac{\pi }{7}} e^{-2 i \sigma_1} \sin ^2(\beta ) \cos (2 \sigma_2)
    \\
    \mathbb{DE}^{Q\pm iU}_{43}
    &=
    \frac{1}{3} \sqrt{\frac{\pi }{2}} e^{-2 i \sigma_1} \sin (\beta ) (\cos (\beta ) \cos (2 \sigma_2)-i \sin (2 \sigma_2))
    \\
    \mathbb{DE}^{Q\pm iU}_{44}
    &=
    \frac{1}{12} \sqrt{\pi } e^{-2 i \sigma_1} ((\cos (2 \beta )+3) \cos (2 \sigma_2)-4 i \cos (\beta ) \sin (2 \sigma_2))
\end{align*}

\ew

\newcommand{\Authname}[2]{#2 #1} 
\newcommand{\etal}{{\it et al.}}
\newcommand{\LSC}{\Authname{Abbott}{B.~P.} \etal\, (LIGO Scientific Collaboration)}
\newcommand{\LVC}{\Authname{Abbott}{B.~P.} \etal\, (LIGO Scientific and Virgo Collaboration)}
\newcommand{\LVK}{\Authname{Abbott}{R.} \etal\, (LIGO Scientific, Virgo, and KAGRA Collaborations)}

\newcommand{\Title}[1]{}               

\newcommand{\arxiv}[1]{\href{http://arxiv.org/abs/#1}{{arXiv:}#1}}
\newcommand{\PRD}[3]{\href{https://doi.org/10.1103/PhysRevD.#1.#2}{{Phys. Rev. D} {\bf #1}, #2 (#3)}}
\newcommand{\PRDR}[3]{\href{https://doi.org/10.1103/PhysRevD.#1.#2}{{Phys. Rev. D} {\bf #1}, #2(R) (#3)}}
\newcommand{\PRL}[3]{\href{https://doi.org/10.1103/PhysRevLett.#1.#2}{{Phys. Rev. Lett.} {\bf #1}, #2 (#3)}}

\newcommand{\MNRAS}[4]{\href{https://doi.org/10.1093/mnras/#1.#4.#2}{{Mon. Not. R. Astron. Soc.} {\bf #1}, #2 (#3)}}
\newcommand{\CQGii}[5]{\href{https://doi.org/10.1088/0264-9381/#1/#4/#5}{{Classical Quant. Grav.} {\bf #1}, #2 (#3)}}
\newcommand{\CQG}[4]{\href{https://doi.org/10.1088/1361-6382/#4}{{Class. Quant. Grav.} {\bf #1}, #2 (#3)}}
\newcommand{\JCAP}[3]{\href{https://doi.org/10.1088/1475-7516/#3/#1/#2}{{J. Cosmol. Astropart. Phys.} #1 (#3) #2}}

\newcommand{\LRR}[4]{\href{https://doi.org/10.1007/#4}{{Liv. Rev. Rel.} {\bf #1}, #2 (#3)}}
\newcommand{\ApJ}[4]{\href{https://doi.org/#4}{{Astrophys. J.} {\bf #1}, #2 (#3)}}


\raggedright

\end{document}